\documentclass[a4paper,11pt]{article}
\pdfoutput=1 \usepackage{jheppub} \usepackage[T1]{fontenc} \usepackage{bm}
\usepackage{braket}

\usepackage[finalizecache=false, frozencache=true, cachedir=./]{minted}  

\usepackage[utf8]{inputenc}
\DeclareUnicodeCharacter{0394}{\ensuremath{\Delta}}

\title{Analytic reconstruction with massive particles: one-loop amplitudes for $0 \to \bar{q}qt\bar{t}H$}
\author[a]{John M. Campbell,}
\emailAdd{johnmc@fnal.gov}
\author[b]{Giuseppe De Laurentis,}
\emailAdd{giuseppe.delaurentis@ed.ac.uk}
\author[c]{R. Keith Ellis,}
\emailAdd{keith.ellis@durham.ac.uk}

\affiliation[a]{Fermilab, PO Box 500, Batavia IL 60510-5011, USA}
\affiliation[b]{Higgs Centre for Theoretical Physics, University of Edinburgh, Edinburgh, EH9 3FD, UK}
\affiliation[c]{Institute for Particle Physics Phenomenology, Durham University, Durham, DH1 3LE, UK}

\preprint{FERMILAB-PUB-25-0255-T,\, IPPP/25/21} \abstract{

We present an analytic reconstruction of one-loop amplitudes for the
process $0 \to \bar{q}qt\bar{t}H$. Our calculation is a novel use of
analytic reconstruction, retaining explicit covariance in the massive
spin states through the massive spinor-helicity formalism.  The
analytic reconstruction relies on embedding the massive five-point kinematics in a
fully massless eight-point phase space while still building a
minimal ansatz directly in the five-point phase space.
In order to obtain compact analytic expressions it is necessary to
identify suitable partial fraction decompositions and extract common
numerator factors, which we achieve through careful inspection of
limits in which pairs of denominators vanish.  We find that
the resulting amplitudes are more numerically efficient than
ones computed using automatic methods but that the gains are not as
significant as in the massless case, at least at present. The
method opens the door to applications at two-loop order, where
numerical efficiency and improvements in the reconstruction
methodology are more crucial, especially with regards to the number
of free parameters in the ansatz.}

\def\cg{c_\Gamma}
\def\lf{{\rm lf}}
\def\boldsigma{{\bm{\sigma}}}
\newcommand{\da}{{\dot{\alpha}}}
\newcommand{\db}{{\dot{\beta}}}
\def\Gramudxtxq{\Delta_{12|3|4|5}}
\def\Gramudxtq{\Delta_{12|34|5}}
\def\Gramudxt{\Delta_{12|3|45}}

\def\st{\tilde{s}}
\def\e{\epsilon}
\def\cG{r_{\Gamma}}

\def\x{{\times}}

\def\three{{\bf 3}}
\def\four{{\bf 4}}
\def\five{{\bf 5}}

\def\beq{\begin{equation}}
\def\eeq{\end{equation}}
\def\beqn{\begin{eqnarray}}
\def\eeqn{\end{eqnarray}}
\def\nn{\nonumber}

\def\spa#1.#2{\left\langle#1\,#2\right\rangle}
\def\spb#1.#2{\left[#1\,#2\right]}
\def\spaa#1.#2.#3{\langle\mskip-1mu{#1} 
                  | #2 | {#3}\mskip-1mu\rangle}
\def\spbb#1.#2.#3{[\mskip-1mu{#1}
                  | #2 | {#3}\mskip-1mu]}
\def\spab#1.#2.#3{\langle\mskip-1mu{#1} 
                  | #2 | {#3}\mskip-1mu]}
\def\spba#1.#2.#3{[\mskip-1mu{#1} 
                  | #2 | {#3}\mskip-1mu\rangle}
\def\spaba#1.#2.#3.#4{\langle\mskip-1mu{#1} 
                  | #2 | #3 | {#4}\mskip-1mu\rangle}
\def\spbab#1.#2.#3.#4{[\mskip-1mu{#1} 
                  | #2 | #3 | {#4}\mskip-1mu]}
\def\spabab#1.#2.#3.#4.#5{\langle\mskip-1mu{#1}
                  | #2 | #3 | {#4}| {#5} \mskip-1mu]}
\def\spbaba#1.#2.#3.#4.#5{[\mskip-1mu{#1} 
                  | #2 | #3 | {#4}| {#5}\mskip-1mu\rangle}
\def\tr#1.#2{\text{tr}(#1|#2)}
\def\qb{\bar{q}}
\def\Qb{\bar{Q}}
\def\cA{{\cal A}}
\def\slsh{\rlap{$\;\!\!\not$}}     \def\three{{\bf 3}}
\def\four{{\bf 4}}
\def\five{{\bf 5}}

\begin{document} 
\maketitle
\flushbottom

\section{Introduction}
\label{sec:intro}
A strong effort is underway to produce theoretical predictions for LHC processes of a precision commensurate
with the expected experimental precision at the conclusion of the high luminosity
LHC. For many processes this will require at least evaluation at next-to-next-to leading order in the strong coupling,
necessitating the calculation of two-loop amplitudes for processes with more than four external legs. This requires analytic
results which are numerically fast and stable to evaluate. This can be achieved by expressing the amplitudes in
compact partial fraction form. Much progress in obtaining results for amplitudes in this form has been facilitated
by the advent of analytic reconstruction
techniques~\cite{Peraro:2016wsq,vonManteuffel:2014ixa,Abreu:2017xsl,Badger:2018enw,Abreu:2018zmy,Laurentis:2019bjh}.

For the most part these reconstruction techniques have focussed on the reconstruction of amplitudes
containing light fermions (which at high energy can be considered effectively massless) in combination
with massless gluons and electroweak bosons. However some of the most interesting processes at high energy
contain top quarks, which cannot be taken to be massless. In this paper we perform an exploratory application
of these techniques to processes involving massive top quarks. We have chosen to consider the subprocess,
$0 \to \qb(p_1)+q(p_2)+Q(p_3)+\Qb(p_4)+H(p_5)$ evaluated at one-loop order. This process is one of the subprocesses
contributing to $t \bar{t}H$ production at the LHC.

The $\bar{q}qt\bar{t}H$ amplitude at one-loop order has first been considered more than twenty years ago,
as part of the next-to-leading order calculation of the $t \bar{t}H$ 
production process~\cite{Reina:2001bc,Dawson:2002tg,Beenakker:2001rj,Beenakker:2002nc}.
Progress aimed at the calculation of the two loop virtual corrections to the $t \bar{t}H$ process has been
reported in refs.~\cite{Catani:2021cbl,Chen:2022nxt,FebresCordero:2023pww,Buccioni:2023okz}.
The implications for phenomenology of the NLO calculations and the more recent approximate
NNLO calculations~\cite{Catani:2022mfv,Devoto:2024nhl,Balsach:2025jcw}
are not the main thrust of this paper. For details on that subject we refer
the reader to the references cited.

\section{Reconstruction of massive spinor-helicity amplitudes}
\label{Reconstruction}
The main objective of the present work is to adapt the technology for
analytic reconstruction in redundant variables
\cite{Laurentis:2019bjh, DeLaurentis:2022otd} to loop amplitudes
involving external massive fermions by demonstrating they can be
reconstructed in the massive spinor-helicity formalism
\cite{Arkani-Hamed:2017jhn, Ochirov:2018uyq, Shadmi:2018xan}. This is
the last remaining kind of external field left to consider for
amplitudes in the Standard Model; amplitudes with a single
\cite{Budge:2020oyl} and multiple \cite{Campbell:2024tqg} massive
scalars and a single \cite{DeLaurentis:2025dxw} and multiple
\cite{Campbell:2022qpq} massive vectors have been previously
reconstructed. Two of these most recent computations
\cite{Campbell:2024tqg, DeLaurentis:2025dxw} have relied on minimal
parametrizations tailored to the specific kinematics under study. For
the case of massive external fermions, we consider one-loop
corrections to the five-point process
\begin{equation}\label{eq:process}
  0 \to \qb(p_1)+q(p_2)+Q(p_3)+\Qb(p_4)+H(p_5)\, ,
\end{equation}
with a pair of light $q\bar q$ and a pair of heavy $Q\bar Q$
quarks. The concurrent presence of both two massive fermions and a
massive scalar provides a stringent test of the technology.

The dependence of the amplitude on covariant tensors transforming
under the $\text{SL}(2, \mathbb{C})$ Lorentz group and the
$SO(3)\equiv SU(2)$ spin group is easily stated. The tensors are
\begin{equation}\label{eq:covariant-tensors}
  |1\rangle_{\alpha}, \; [1|_{\da}, \; |2\rangle_{\alpha}, \;
    [2|_{\da}, \; |\three^I\rangle_{\alpha}, \;
      [\three^I|_{\da}, \; |\four_J\rangle_{\alpha}, \;
        [\four_J|_{\da}, \; \five_{\alpha\da}  \; , 
\end{equation}
where we have made explicit the $\text{SL}(2, \mathbb{C})$ indices
$\alpha$ and $\da$, and the $SU(2)$ indices $I$ and $J$.
We shall refer to objects such as $|\three^I\rangle_{\alpha}$ as
spin-spinors~\cite{Arkani-Hamed:2017jhn}.
As a reminder of their massive nature, $\three,\four$ and $\five$ are indicated by boldface
numbers.
Further information on the spin-spinor formalism is given in Appendix~\ref{Spinor_techniques}.
In amplitudes, all $\text{SL}(2, \mathbb{C})$ indices will be contracted,
but the $SU(2)$ ones will not be. The $\text{SL}(2, \mathbb{C})$ indices
can often be suppressed, since their contraction rules are mandated by the
position of the square and angle brackets.
The $SU(2)$ indices, when suppressed, are
implied to be as in eq.~\eqref{eq:covariant-tensors}. The rank-two
spinors for the massive fermions can be built as
\begin{gather}
  |\three^I\rangle_{\alpha}[\three_I|_{\da} =
    \three_{\alpha\da} \, , \\
    |\four^J\rangle_{\alpha}[\four_J|_{\da} =
      \four_{\alpha\da} \, ,
\end{gather}
where a summation over $I$ and $J$ is implied, and where the $SU(2)$
indices are raised and lowered through contraction with the totally
anti-symmetric Levi-Civita tensor
$\epsilon^{IJ}=-\epsilon_{IJ}=i\sigma_2$ on the right\footnote{For the definition of
the epsilon tensor, see also eq.~\eqref{eq:epdef}.}. Momentum
conservation imposes a redundancy in these variables
\begin{equation}
  |1\rangle[1|+|2\rangle[2|+|\three^I\rangle[\three_I|+|\four^J\rangle[\four_J|
          + \five = 0 \, .
\end{equation}
In practice, we often lift this redundancy by eliminating $\five$ in
favour of the other variables.

It is important, especially when performing numerical manipulations,
for the spin indices to be uniformly raised or lowered for each
massive external fermion. This allows for simplifications through
applications of the Dirac equation
\begin{eqnarray}\label{eq:DiracEq}
  \langle \three^I|^\alpha \three_{\alpha\da} &=&   + [\three^I|_{\da} \, m_3\, ,\,\;\;\;
  [\three^I|_\da \three^{\da\alpha} =  +\langle\three^I|^{\alpha} \, \bar m_3 \, , \\
  \three^{\da\alpha} |\three^I\rangle_\alpha &=& - m_3 \, |\three^I]^{\da}\, , \,\;\;\;
    \three_{\alpha\da} |\three^I]^{\da} = - \bar m_3 \, |\three^I\rangle_\alpha \, ,
\end{eqnarray}
where (see also ref.~\cite[Appendix A]{Shadmi:2018xan})
\begin{equation}
  m_3 \epsilon^{JI} = \langle \three^{I} | \three^{J} \rangle \, ,\;\;\; \bar m_3 \epsilon^{IJ} = [ \three^{I} | \three^{J} ] \, ,
\end{equation}
and equivalently with the $I,J$ indices lowered, such that
\begin{gather}
  m_3^2 = m_3 \bar m_3 = \langle \three^{I=2} |
  \three^{I=1} \rangle [ \three_{I=1} | \three_{I=2} ] = p_\three^2 \,
  .
\end{gather}
The most generic kinematic configuration has all four masses $m_3$,
$\bar m_3$, $m_4$, $\bar m_4$ distinct. However, we impose additional
constraints forcing all these masses to be equal. This amounts to three
additional redundancies in the space of
eq.~\eqref{eq:covariant-tensors}, namely
\begin{align}
  \langle \three^{I=1} | \three^{I=2} \rangle + [\three_{I=1} | \three_{I=2}] &= - m_3 + \bar m_3 = 0 \, , \label{eq:mass-constraints-first}\\
  \langle \three^{I=1} | \three^{I=2} \rangle - \langle \four^{J=1} | \four^{J=2} \rangle &=  - m_3 + m_4 = 0 \, , \\
  \langle \four^{J=1} | \four^{J=2} \rangle + [\four_{J=1} | \four_{J=2}] &= - m_4 + \bar m_4 = 0 \, , \label{eq:mass-constraints-last}
\end{align}
which automatically implies the last relation $[\three_{I=1} |
  \three_{I=2}]-[\four_{J=1} | \four_{J=2}]=\bar m_3-\bar
m_\four=0$.

\subsection{Embedding in higher-point massless phase space}

To facilitate relating this case to the more familiar case with only
massless particles, we shall now see how the five-point phase space
discussed so far relates to a fully massless eight-point phase
space. The massive spinor formalism can ultimately be viewed as being
just a neat relabelling of the fully massless case
\cite{Conde:2016vxs, Conde:2016izb, Marzolla:2017ego}. The relation
among four momenta ($p^\mu$), or equivalently among rank-two spinors
($p_{\alpha\dot\alpha}$), reads
\begin{equation}\label{eq:eight-point-to-five-point-first}
  1 \rightarrow 1, \; 2 \rightarrow 2, \; \three \rightarrow 3 + 4, \; \four
  \rightarrow 5 + 6, \; \five \rightarrow 7 + 8 \, ,
\end{equation}
where labels on the left-hand side refer to five-point phase space of
eq.~\eqref{eq:covariant-tensors}, and those on the right-hand side to
the fully massless eight-point one. Following the same convention, the
states for the massive spinors are mapped as follows
\begin{equation}
  |\three^{I=1}\rangle \rightarrow |3\rangle\, ,\;
  [\three_{I=1}| \rightarrow [3|\,,\;
  |\three^{I=2}\rangle \rightarrow |4\rangle\, ,\;
      [\three_{I=2}| \rightarrow [4| \, ,
\end{equation}
and likewise
\begin{equation}\label{eq:eight-point-to-five-point-last}
  |\four^{J=1}\rangle \rightarrow |5\rangle\, ,\;
  [\four_{J=1}| \rightarrow [5|\, ,\;
  |\four^{J=2}\rangle \rightarrow |6\rangle\, ,\;
  [\four_{J=2}| \rightarrow [6| \, .
\end{equation}

In the algebro-geometric language of ref.~\cite{DeLaurentis:2022otd},
any phase space point living on the variety associated
to\footnote{$S_8$ is the polynomial ring in 8 massless spinor pairs,
see ref.~\cite{DeLaurentis:2022otd}.}
\begin{equation}\label{eq:q-ring-variety}
  \Big\langle \sum_{i=1}^{8} |i\rangle[i|, \langle 34\rangle + [34],
    \langle 34\rangle - \langle 56\rangle, \langle 56\rangle + [56],
    \Big\rangle_{S_8} \, ,
\end{equation}
is a valid phase space point for our calculation, including momentum
conservation and the constraints for the massive quarks of
eqs.~\eqref{eq:mass-constraints-first}-\eqref{eq:mass-constraints-last}. This
reformulation gives us direct access to finite field and
$p\kern0.2mm$-adic phase space points within the usual massless
setup. We generate an arbitrary phase space point with eight massless
legs that lives on the variety of eq.~\eqref{eq:q-ring-variety} and
then map it down to five-point phase space through the inverse map of
eqs.~\eqref{eq:eight-point-to-five-point-first}-\eqref{eq:eight-point-to-five-point-last}.
In appendix \ref{lips}, we provide example code to perform these
operations as implemented in the Python package \textsc{Lips}
\cite{DeLaurentis:2023qhd}.

\subsection{Massive spinor-helicity parametrization of numerators}

While the eight-point embedding works well for the numerical and
semi-numerical aspects of the calculation discussed so far, a naive
analytic eight-point parametrization of the numerators would severely
over-parameterize the space of allowed monomials. For instance, no
rank-one spinor for legs $7$ and $8$ can appear while those for legs
$3$, $4$, $5$, and $6$ have a degree bound of one, unless contracted
and summed over to form the rank-two spinors
$\three_{\alpha\da}$, $\four_{\alpha\da}$ and
$\five_{\alpha\da}$. For this reason it is best to build the
parametrization (or \textit{ansatz}) directly in five-point phase
space.

Helicity amplitudes are covariant in the spin indices $I$ and $J$, and
since we are manifesting such covariance it will be sufficient to
determine the amplitude for a single choice of $I$ and $J$, say
$I=J=1$. We can thus drop the $I=2$ and $J=2$ variables, at the price
of explicitly reintroducing the corresponding rank-two spinors. We
define the set of variables,
\begin{equation}\label{eq:covariant-space-ansatz}
  \underline X = \big\{ |1\rangle, [1|, |2\rangle, [2|, |\three^{I=1}\rangle,
      [\three^{I=1}|, \three, |\four_{J=1}\rangle, [\four_{J=1}|,
          \four \big\} \, .
\end{equation}
We must reintroduce $\three$ and $\four$ because the absence of the
$I=J=2$ rank-one components does not allow us to otherwise obtain
them.  This approach has an important benefit, namely that we can
manifest the degree bound of 1 for the states of the massive
fermions. We stress that it is not required to perform the
reconstruction for all choices of $I$ and $J$. A single choice is
sufficient to fully fix all four.  This is guaranteed as long as no
special choice is made for the decomposition in eq.~\eqref{eq:eight-point-to-five-point-first},
which we ensure by choosing arbitrary directions for two of the massless
momenta, $p_3$ and $p_5$.

Since numerators are Lorentz invariant, we need to convert from the
covariant space of eq.~\eqref{eq:covariant-space-ansatz} to a space of
invariant spinor brackets. We identify the following irreducible
contractions as a sufficient set to span the Lorentz invariant space,
\begin{equation}
  \begin{gathered}\label{eq:invariant-space-ansatz}
    \underline{\mathcal{X}} = \big\{
\spa1.2,\, \spa\three.1,\, \spa1.\four,\, \spa\three.2,\, \spa2.\four,
    \spa\three.\four,\, \\
\spb1.2,\, \spb\three.1,\, \spb1.\four,\, \spb\three.2,\, \spb2.\four,
    \spb\three.\four,\, \\
\spab1.\three.1,\, \spab1.\three.2,\, \spab1.\three.\four,
    \spab2.\three.1,\, \spab2.\three.2,\, \spab2.\three.\four,
    \spba1.\three.\four,\, \spba2.\three.\four,\, \\
\spab1.\four.1,\, \spab1.\four.2,\, \spba\three.\four.1,
    \spab2.\four.1,\, \spab2.\four.2,\, \spba\three.\four.2,
    \spab\three.\four.1,\, \spab\three.\four.2,\, \\
\spaba1.\three.\four.1,\, \spaba1.\three.\four.2,\, \spaba2.\three.\four.2,
    \spbab1.\three.\four.1,\, \spbab1.\three.\four.2,\, \spbab2.\three.\four.2,\, \\
\tr\three.\three,\, \tr\three.\four,\, \tr\four.\four,\, m^2
\big\} \, ,
  \end{gathered}
\end{equation}
where we have made manifest the origin of $\langle\three|$,
$[\three|$, $|\four\rangle$, and $|\four]$ from four-component Dirac
spinors by having the former two always on the left and the latter two
always on the right.
We obtained this set by using momentum conservation to explicitly
eliminate $\five$ and the Dirac equation eq.~\eqref{eq:DiracEq} to reduce
contractions of the form $\langle\three|\three|$, $[\three|\three|$,
  $|\four|\four\rangle$, and $|\four|\four]$. Furthermore, the two
traces $\tr\three.\three$ and $\tr\four.\four$ can be eliminated in
favour of $m^2$, due to the additional on-shell constraints, while
the last remaining trace $\tr\three.\four$ is equal to $2 p_3 \cdot
p_4$. More generally, we define the following Mandelstam-like
invariants,
\beqn s_{ij}&=&(p_i+p_j)^2\, ,\;\;\;\; s_{ijk}=(p_i+p_j+p_k)^2\, , \nn \\
\tilde{s}_{ij}&=&2 p_i\cdot p_j\,, \;\;\;\;\tilde{s}_{ijk}=2 p_i\cdot p_j+2 p_j\cdot p_k+2 p_k\cdot p_i\, ,
\eeqn
so that we can also write spinor quantities in terms of these variables, e.g.
$\spab1.\three.1 = 2p_1 \cdot p_3 = \tilde s_{13}$. The list of independent invariants of
eq.~\eqref{eq:invariant-space-ansatz} does not contain a linear mass,
because the $m$ dependence can be fixed by the accompanying
combination of angle and square brackets for the massive fermion
states.

In the presence of massive fermions, amplitudes lose uniform scaling
under the respective little group transformations. Therefore, we cannot
assign a little group weight to legs 3 and 4 (labelled below in eq.~\eqref{eq:massive-ansatz} as
``not a number'', NaN). Nevertheless, we can still decompose the
complete ansatz into a sum of four ans\"atze with uniform scaling,
\begin{gather} 
  \text{ansatz}(d, \{w_1, w_2, \text{NaN}, \text{NaN}, 0\}) = \label{eq:massive-ansatz}\\ \begin{array}{@{\hspace{-20pt}}c@{\hspace{4pt}}c@{\hspace{4pt}}l@{\hspace{4pt}}c@{\hspace{4pt}}c@{\hspace{4pt}}c@{\hspace{4pt}}l@{}}
    \qquad (1 - \delta) & \times & \text{ansatz}(d, \{w_1, w_2, 1, 1, 0\}) & + & \delta & \times & m \;\, \text{ansatz}(d - 1, \{w_1, w_2, 1, 1, 0\}) \\
    \qquad (1 - \delta) & \times & \text{ansatz}(d, \{w_1, w_2, -1, -1, 0\}) & + & \delta & \times & m \;\, \text{ansatz}(d - 1, \{w_1, w_2, -1, -1, 0\}) \\
    \qquad \delta & \times & \text{ansatz}(d, \{w_1, w_2, 1, -1, 0\}) & + & (1 - \delta) & \times & m \;\, \text{ansatz}(d - 1, \{w_1, w_2, 1, -1, 0\}) \\
    \qquad \delta & \times & \text{ansatz}(d, \{w_1, w_2, -1, 1, 0\}) & + & (1 - \delta) & \times & m \;\, \text{ansatz}(d - 1, \{w_1, w_2, -1, 1, 0\}) \, .
  \end{array} \nonumber
\end{gather}
In eq.~\eqref{eq:massive-ansatz}, $d$ is the mass dimension and
$w_1,w_2$ are the phase weights of the light fermions in the ansatz.
Moreover, in the above, we have made explicit that each uniform-weight
ansatz can either appear with no linear $m$ dependence, or with a
factor of $m$ which must lower the mass dimension by one
unit. This dependence, encoded in $\delta$, depends on the mass
dimension $d$, and the phase weights $w_1$, and $w_2$ through,
\begin{equation}
  \delta = \begin{cases}
    0 & \text{if } \frac{1}{2}(w_1+w_2+2) \bmod 2 = d \bmod 2, \\
    1 & \text{otherwise}.
  \end{cases}
\end{equation}

The problem is now to construct each of the ans\"atze in the right
hand side of eq.~\eqref{eq:massive-ansatz}, that is to enumerate a
complete set of independent monomials in the variables of
eq.~\eqref{eq:invariant-space-ansatz} with given mass dimension and
little group weights. We follow the procedure described in
ref.~\cite[Section 2.2]{DeLaurentis:2022otd}, barring a couple of
caveats that we now discuss. For the present computation, the
covariant quotient ring is taken to be in the components of the
rank-one and rank-two spinors of
eq.~\eqref{eq:covariant-space-ansatz}, modulo a single constraint
equating the masses of the top quarks,
\begin{equation}
  R = \mathbb{F}\big[\underline X \big] \Big / \big\langle \tr\three.\three-\tr\four.\four \big\rangle_{\mathbb{F}[\underline X ]} \, ,
\end{equation}
where the subscript denotes in which ring the ideal is taken and
$\mathbb{F}$ is a generic number field. The invariant polynomial ring
is in the brackets listed in
eq.~\eqref{eq:invariant-space-ansatz}. The corresponding invariant
quotient ring,
\begin{equation}
  \mathcal{R} = \mathbb{F}\big[ \underline{\mathcal{X}} \big] \Big /  \big\langle \tr\three.\three-\tr\four.\four,\,\tr\three.\three-2m^2,\;\text{Schouten identities}\big\rangle_{\mathbb{F}[ \underline{\mathcal{X}}] } \, ,
\end{equation}
can, in principle, be obtained by elimination of variables from the
covariant one. This requires obtaining a Gr\"obner basis with a
monomial block ordering in the ring with both covariant and invariant
variables for the ideal generated by the mass constraint. It would
yield the entire set of constraints (Schouten identities) in the
invariant ring. Unfortunately, it is unclear how long such a Gr\"obner
basis computation would take. We therefore decided to split the problem
into two easier problems. First, we consider the case where the top quarks
are taken to be scalars. That is, we remove all variables associated
with the top states, both from the covariant ring,
\begin{gather}
  S_{\text{scalar-tops}} = \mathbb{F}\big[ |1\rangle, [1|, |2\rangle, [2|, \three, \four
         \big] \\
  R_{\text{scalar-tops}} = S_{\text{scalar-tops}} \Big / \big\langle \tr\three.\three-\tr\four.\four
      \big\rangle_{S_{\text{scalar-tops}}}\, ,
\end{gather}
and the invariant ring,
\begin{gather}
  \underline{\mathcal{X}}_{\text{scalar-tops}} = \big\{
\spa1.2,\, \spb1.2,\,
  \spab1.\three.1,\, \spab1.\three.2,\,
  \spab2.\three.1,\, \spab2.\three.2,\,
  \spab1.\four.1,\, \spab1.\four.2,\, 
  \spab2.\four.1,\, \spab2.\four.2,\, \nonumber \\
\qquad\qquad \spaba1.\three.\four.1,\, \spaba1.\three.\four.2,\, \spaba2.\three.\four.2,
  \spbab1.\three.\four.1,\, \spbab1.\three.\four.2,\, \spbab2.\three.\four.2,\, \nonumber \\
  \qquad\qquad \tr\three.\three,\, \tr\three.\four,\, \tr\four.\four,\, m^2
\big\} \, , \\
  \mathcal{S}_{\text{scalar-tops}} = \mathbb{F} \big[ \underline{\mathcal{X}}_{\text{scalar-tops}} \big] \, ,\\
  \mathcal{R}_{\text{scalar-tops}} = \mathcal{S}_{\text{scalar-tops}} \, \big/ \, \mathcal{J}_{\text{scalar-tops}} \, .
\end{gather}
We compute the ideal $\mathcal{J}_{\text{scalar-tops}}$ that encodes the
redundancies, including the trivial rewriting $\tr\three.\three=2m^2$,
by the elimination algorithm. We obtain more than 90 generators for
the ideal $\mathcal{J}_{\text{scalar-tops}}$, which has dimension $9$ in
the space of $20$ variables of
$\underline{\mathcal{X}}_{\text{scalar-tops}}$.

Finally, we supplement these redundancies with (some) Schouten
identities involving the top states. These can be obtained numerically
(by looking at the null-space of a redundant ansatz), or
analytically. Unfortunately, even with this optimization, which could
be thought of as a block ordering on the variables
$\underline{\mathcal{X}}$, we are unable to obtain a complete
Gr\"obner basis. This means that we are unable to fully remove
redundancies from the ansatz during its construction. However, given
that the top spinor states cannot appear with degree higher than one,
we find that the impact of the leftover redundancies is negligible, as they
can be easily removed numerically before or during the ansatz
fitting. In the end, the fitted ansatz is minimal for this process.

\subsection{Primary decompositions for partial fractions and numerator factors}
\label{subsec:primdec}

For the purposes of the analytic reconstruction, the most important
polynomials are those appearing as irreducible denominator
factors. In the present computation, we identify the following factors,
\begin{equation}\label{eq:irreducible-denominator-factors}
  \begin{gathered}
    \underline{\mathcal{D}} = \big\{
\spa1.2,\, \spb1.2,\, \\
s_{123},\, s_{124},\, s_{34},\, (s_{123}-m^2),\,
    (s_{124}-m^2),\,(s_{34}-4m^2),\,\\
\spab1.\three.1,\, \spab1.\three.2,\, \spab1.\four.1,\, \spab1.\four.2,\,
    \spab1.(2+\three).1,\, \spab2.(1+\three).2,\, \\
\spaba1.\four.\five.1,\,\spaba1.\three.\five.1,\,\spbab2.\five.\three.2,\,\spbab2.\five.\four.2,\,\\
\spabab1.\three.(1+2).\four.2,\, \spabab1.\four.(1+2).\three.2,\, \spabab2.\three.(1+2).\four.1,\, \spabab2.\four.(1+2).\three.1,\, \\
\Delta_{12|34|5},\, \Delta_{123|4|5},\, \Delta_{124|3|5},\,
    \Delta_{12|3|45},\,  \Delta_{12|4|35},\, \\
\Delta_{12|3|4|5},
\big\} \, .
  \end{gathered}
\end{equation}
This set is closed under both a $1\leftrightarrow 2$ and a
$3\leftrightarrow 4$ swap. Several of these poles are spurious in the
amplitude but not in the individual coefficients. In some cases, we
use a $\spab1.\five.2$ spurious pole to allow for extra partial
fraction decompositions. This allows us to limit the ansatz size without
complicating the computation.

Allowed partial fraction decompositions as well as common numerator
factors are controlled by the behaviour of functions near regions of
phase space where pairs of these denominator factors vanish
\cite{Laurentis:2019bjh, DeLaurentis:2022otd}.  As a simple concrete
example we shall take the case of the coefficient of a triangle
integral to be presented later in eq.~\eqref{eq:c13x24m0m}.  This
function has two simple poles, $\spaba1.\three.\five.1$ and
$\spbab2.\five.\four.2$. A natural question is then whether we can
write
\beq\label{eq:tentative-decomposition}
c_{13\x24m0m} = \frac{\mathcal{N}}{\spaba1.\three.\five.1\spbab2.\five.\four.2} \stackrel{?}{=} \frac{\mathcal{N}_1}{\spaba1.\three.\five.1}+\frac{\mathcal{N}_2}{\spbab2.\five.\four.2} \, ,
\eeq
for some $\mathcal{N}_1$ and $\mathcal{N}_2$. To answer this question,
we need to check how $\mathcal{N}$ behaves when
$\spaba1.\three.\five.1$ and $\spbab2.\five.\four.2$
vanish. Generating random phase space points such that these two
invariants are small, we see that $c_{13\x24m0m}$ displays two
behaviours, depending on the point chosen. In some cases, it has a
double pole, in some cases it has a simple pole. If it only had a
simple pole, then the decomposition of
eq.~\eqref{eq:tentative-decomposition} would hold due to Hilbert's
Nullstellensatz\footnote{Assuming the two polynomials generate an
ideal that is radical.}. Since in some cases it has a double pole, then
there exist no $\mathcal{N}_1$, $\mathcal{N}_2$ such that the
decomposition holds.

In order to quantify where each behaviour happens, a \textit{primary
  decomposition} of the corresponding ideal is required. Computing
such decompositions can be highly non-trivial. In this case, there are
two associated primes,
\begin{align}\label{eq:decomp_spaba1351_spbab2542}
\big\langle \spaba1.\three.\five.1,\, \spbab2.\five.\four.2 \big\rangle = \; &\big\langle \,  \spab1.\three.2,\, \spab1.\four.2,\, \spaba1.\three.\five.1,\, \spbab2.\five.\four.2
\, \big\rangle\; \cap \\
&\big\langle \, \spaba1.\three.\five.1,\, \spbab2.\five.\four.2, |\five|2]\langle1|\three| - |1+\three|2]\langle1|\five| \, \big\rangle \;, \nonumber
\end{align}
where the second prime ideal has a generator with two open
indices. These ideals are symmetric under $\{1\leftrightarrow 2,
\three \leftrightarrow \four, \langle\rangle \leftrightarrow []\}$. We
will refer to varieties associated with prime ideals as
\textit{branches}. These are the regions of phase space where the
generators of the corresponding ideal vanish. The decomposition of
eq.~\eqref{eq:decomp_spaba1351_spbab2542} is reminiscent of that given
in ref.~\cite[Appendix C]{Campbell:2024tqg}. Indeed, a similar
relation exists which elucidates the decomposition, \beq
\begin{gathered}
  |\five|2]\spaba1.\three.\five.1[2| + |1\rangle\spbab2.\five.\four.2\langle1|\five| = \spab1.\five.2 \Big( |\five|2]\langle1|\three| - |1+\three|2]\langle1|\five| \Big) \, ,
\end{gathered}
\eeq
where the open-index expression in the parenthesis is the same as in
the second prime ideal of eq.~\eqref{eq:decomp_spaba1351_spbab2542}
and the factor $\spab1.\five.2$ belongs to the first prime
ideal. Similar relations should exist for $\spab1.\three.2$ and
$\spab1.\four.2$. Such an identity demonstrates that the ideal generated
by just the two denominator factors is not prime, since there exists a
polynomial combination which factorizes as a product of two terms
neither of which is in the ideal.

Going back to considering $c_{13\x24m0m}$, we can now state that it is
on the first branch that the function has a double pole, while on the
second branch it has a simple pole. We can then allow for a partial
fraction decomposition by adding as a spurious pole a polynomial that
belongs to the first prime ideal, since this will ensure that the new
numerator vanishes on both branches. Due to the symmetry of this
function, we choose $\spab1.\five.2$. We conclude that the following
representation exists
\beq\label{eq:allowed-decomposition}
c_{13\x24m0m} =
\frac{\mathcal{N}'_1}{\spab1.\five.2\spaba1.\three.\five.1}+\frac{\mathcal{N}'_2}{\spab1.\five.2\spbab2.\five.\four.2}
\, ,
\eeq
for some new polynomials $\mathcal{N}'_1$, $\mathcal{N}'_2$ to be
determined. This would allow us to use
eq.~\eqref{eq:decomp_spaba1351_spbab2542} to constrain the analytic
form of this function, even without knowing an explicit covariant
representation for the generator of the second branch. On the other
hand, if we wanted to avoid the insertion of the spurious pole, while
still constraining the analytic form, we could use the fact that the
numerator $\mathcal{N}$ in the least common denominator form must be a
contraction of the generator with two open indices of the second
branch.

Let us now see how similar reasoning allows us to constrain the
numerators $\mathcal{N}'_1$, $\mathcal{N}'_2$ by identifying one of
their factors. In ref.~\cite{Campbell:2022qpq} we provided (as a
conjecture) the decomposition
\beq
\sqrt{\big\langle \spaba7.3+4.5+6.7\,, \Delta_{712|34|56} \big\rangle} = \big\langle \Delta_{712|34|56},\, |3+4|5+6|7\rangle-|5+6|3+4|7\rangle \big\rangle \, ,
\eeq
in massless seven point phase space. With the latest developments in
the publicly available Python code \textsc{Syngular}
\cite{syngular} we can now prove this decomposition\footnote{The main
improvement is a semi-numerical calculation of the ideal dimension
through the generation of a finite field point on the associate
variety, setting \texttt{seminumerical\_dim\_computation} in the
primality test.}. One way to recast it to the phase space of the
current computation is
\beq
\sqrt{\big\langle \spbab2.\five.\four.2\,, \Delta_{24|13|5} \big\rangle} = \big\langle \Delta_{24|13|5},\, |2+\four|\five|2]-|\five|2+\four|2] \big\rangle \, .
\eeq
In the asymmetric limit \cite{Campbell:2022qpq} where
$\Delta_{24|13|5}$ and $\spbab2.\five.\four.2$ vanish to second order,
but the difference of open index spinor chains goes to zero linearly,
we observe that $c_{13\x24m0m}$'s pole order drops. Therefore, we
identify $|2+\four|\five|2]-|\five|2+\four|2]$ as a numerator factor
  of the term with $\spbab2.\five.\four.2$ in the denominator.  While
  the contraction in the numerator of eq.~\eqref{eq:c13x24m0m} has
  been expanded, we can make see that it originates from the
  following construction
\beq
\mathcal{N}'_2 \propto ([2|\five|2+\four|-[2|2+\four|\five|) \cdot (|2]\spab\three.1+\three.\four+|\four]\spab\three.\five.2)) \, ,
\eeq
where the proportionality factor is a number and the latter covariant
factor was obtained by fitting an ansatz. Note that we have reversed the spinor
chain to retain the convention of $|\four]$ being contracted from the
  right.

Similar decompositions are instrumental in keeping the complexity under
control, especially in the presence of higher degree polynomials. The
highest degree denominator factor that we encounter is the four-mass
box Gram determinant. It is defined as ($s_3=m^2$)
\beq \label{eq:boxgram}
\Delta_{12|3|4|5} =
\left[-s_{12}s_{34}(s_{34}-4s_3)
  -s_3\tr{1+2}.{\three+\four}^2+\tr{1+2}.\three\tr{1+2}.\four s_{34} \right] /4 \, .
\eeq
We also encounter the three-mass triangle Gram determinants,
\beqn \label{eq:trigram}
\Delta_{12|34|5} &=& \tr{1+2}.{\three+\four}^2/4 - s_{12}s_{34}\, , \nn \\
\Delta_{12|3|45} &=& \tr{1+2}.\three^2/4 - s_{12} s_3\, .
\eeqn
Note that the last two of these have the opposite sign from the customary
definition.

Primary decompositions involving combinations of these Gram determinants
are especially important and
we have used a number of them throughout this computation.
For example, we have
\begin{gather}
  \kern-50mm\big\langle \Delta_{12|3|45},\,\Delta_{12|3|4|5} \big\rangle =
  \big\langle
  \Delta_{12|3|45},\,\Delta_{12|3|4|5},\, \nonumber \\
  \kern30mm (s_{12}\tr\three.\four/2-\tr1+2.\three\tr1+2.\four/4)^2,\, \nonumber \\
  \kern32mm (s_3\tr1+2.\four/2-\tr\three.\four\tr1+2.\three)/4)^2
  \big\rangle \, , \label{eq:decomp_delta12_3_45_and_delta_12_3_4_5}
\end{gather}
which follows directly from
\begin{gather}
s_{12} \Delta_{12|3|4|5} - \Delta_{53|12|4} \Delta_{12|3|45} + (s_{12} \tr\three.\four / 2 - \tr1+2.\three \tr1+2.\four / 4)^2 = 0  \, , \\
s_3 \Delta_{12|3|4|5} - \Delta_{512|3|4} \Delta_{12|3|45} + (s_3 \tr1+2.\four / 2 - \tr\three.\four \tr1+2.\three / 4)^2 = 0 \, .
\end{gather}
The radical of this ideal is obtained simply by removing the squares
\begin{gather}\label{eq:prime_delta12_3_45_delta_12_3_4_5}
  \kern-10mm\sqrt{\big\langle \Delta_{12|3|45},\,\Delta_{12|3|4|5} \big\rangle} =
  \big\langle
  \Delta_{12|3|45},\, (s_{12}\tr\three.\four/2-\tr1+2.\three\tr1+2.\four/4),\, \nonumber \\
  \kern20mm (s_3\tr1+2.\four/2-\tr\three.\four\tr1+2.\three)/4)
  \big\rangle \, .
\end{gather}
The primality test in \textsc{Syngular} shows it to be prime. We can
see this decomposition at play for instance in
eqs.~\eqref{b12xmm}-\eqref{b12xmmGbit}. In particular, the $F$
functions of eqs.~\eqref{b12xmmFbit1}-\eqref{b12xmmFbit2} rewrite the
generators of eq.~\eqref{eq:prime_delta12_3_45_delta_12_3_4_5} with
the $\tilde s$ notation. These factors constrain the numerators of the
terms that have simultaneously both Grams as poles.

Continuing with decompositions between the box and triangle Grams we
have,
\begin{gather}\label{eq:decomp_delta12_34_5_and_delta_12_3_4_5}
  \big\langle \Delta_{12|34|5},\,\Delta_{12|3|4|5} \big\rangle =
  \big\langle
  s_{34},\, \tr1+2.{\three+\four}^2
  \big\rangle \cap
  \big\langle
  \Delta_{12|34|5},\, \tr1+2.{\three-\four}^2 
  \big\rangle \, .
\end{gather}
The first part of the decomposition is clear from the definitions of the
Gram determinants above: both $\Delta_{12|3|4|5}$ and
$\Delta_{12|34|5}$ vanish as
$s_{34} \to 0$ and $\tr{1+2}.{\three+\four}^2 \to 0$.
The second follows from the identity
\beq
4 \Delta_{12|3|4|5} 
 - (s_{34}-4s_3) \Delta_{12|34|5} 
 + s_{34}\tr{1+2}.{\three-\four}^2/4 = 0 \, .
\eeq
The primes in the decomposition are obtained again by removing the
squares. For instance, we can see this decomposition at play in
several terms of eq.~\eqref{eq:c12x34xmmm}. In particular, the
coefficient $c_{12\x34\x mmm}$ has a lower degree of divergence on the
branch with $\tr{1+2}.{\three+\four}$ as a generator than in that with
$\tr{1+2}.{\three-\four}$.

Next, we have a decomposition between a triangle Gram and another denominator factor
\begin{align}\label{eq:decomp_delta12_3_45_and_s123-m2}
  \big\langle \Delta_{12|3|45},\, s_{123}-m^2 \big\rangle =
  \; &\big\langle 
  s_{123}-m^2 ,\, \tr1+2.\three+4m^2 \nonumber 
  \big\rangle\; \cap \\
  \; &\big\langle
  s_{123}-m^2 ,\, \spa1.2
  \big\rangle \, \cap \,
  \big\langle
  s_{123}-m^2 ,\, \spb1.2
  \big\rangle \, .
\end{align}
The last two parts of this decomposition are clear from the definitions.
The first part can be understood from the relation
\beq
4\Delta_{12|3|45}
 - (s_{123}-m^2)\tr{1+2}.\three
 + s_{12}(\tr{1+2}.\three+4m^2) = 0 \, ,
\eeq
For an example use of this
decomposition, see eq.~\eqref{b12xmm}, which diverges faster on the
first branch than the latter two.

A related decomposition is
\begin{equation}\label{eq:decomp_delta123_4_5_and_s123-m2}
  \big\langle \Delta_{123|4|5},\, s_{123}-m^2 \big\rangle =
  \; \big\langle
  s_{123}-m^2 ,\, \tr\five.\four+4m^2
  \big\rangle\; \cap  \; \big\langle 
  s_{123}-m^2 ,\, s_5
  \big\rangle \, ,
\end{equation}
which loses the second branching since leg 5 has non-zero mass.

Lastly, we obtained this decomposition
\begin{align}
  \big\langle \spab1.\three.2, \Delta_{12|3|4|5} \big\rangle =
  \; &\big\langle \spab1.\three.2,\, \spab1.\four.2,\, \spaba1.\three.\five.1,\, \spbab2.\five.\three.2
  \big\rangle\; \cap  \nonumber  \\
  \; &\big\langle |1|2|\three|\four|\three|-|1|\three|1|\four|\three|+|1|\three|\three|\four|1+2|-|\three|2|\three|\four|1+2|
  \big\rangle \, ,
\end{align}
for which the prime ideal generating the second branch has two open
indices, corresponding to four component equations. We used this decomposition in the reconstruction of
$c_{12\x3\x00m}$, which is provided in the ancillary files. We also
provide as an ancillary file a Jupyter notebook,
\texttt{primary-decompositions.ipynb}, to verify these decompositions.

As a closing comment, we note that several of these decompositions are
little-group invariant, such as
eq.~\eqref{eq:decomp_delta12_3_45_and_delta_12_3_4_5} and
\eqref{eq:decomp_delta12_34_5_and_delta_12_3_4_5}, while
e.g.~eq.~\eqref{eq:decomp_delta12_3_45_and_s123-m2} can be made
little-group invariant by intersecting the latter two branches, and
still retains a non trivial decomposition, as in
eq.~\eqref{eq:decomp_delta123_4_5_and_s123-m2}. This shows, as
expected, that in the presence of polynomials of degree higher than
one, even when picking Mandelstam invariants as the reconstruction
variables \cite{Chawdhry:2023yyx}, one needs to account for branching
when the valuations of multiple denominator factors are picked to be
non zero. In particular, the numerators will often have non-trivial
behaviour in limits where denominator factors vanish.

\subsection{Summary of reconstruction through iterated pole subtraction}

We exploit the map to fully massless phase space to recast the
determination of the least common denominators of the integral
coefficients and the generation of singular kinematic configurations
in terms of existing techniques. Explicitly, for each function that we
wish to analytically reconstruct we use the following strategy largely
based on the iterative pole subtraction approach of
ref.~\cite{Laurentis:2019bjh}:

\begin{enumerate}
\item  Use univariate Thiele
interpolation \cite{Peraro:2016wsq} to match the univariate
denominator factors to those of the irreducible polynomials
\cite{Abreu:2018zmy} that are given in
eq.~\eqref{eq:irreducible-denominator-factors} on a univariate slice
that lives on the variety of eq.~\eqref{eq:q-ring-variety} where our
calculation is valid \cite{Campbell:2024tqg}.
This provides a starting point where the structure of the function
is understood in common denominator form.
\item Use $p\kern0.2mm$-adic evaluations near irreducible varieties where
pairs of the denominator factors vanish \cite{DeLaurentis:2022otd}.
Identify valid partial fraction decompositions
(when the numerator vanishes uniformly on all branches) and possible
numerator factors (when the numerator vanishes on certain
branches) using the primary decompositions discussed in section~\ref{subsec:primdec}.
\item Sample the coefficients in limits that isolate terms in the partial
  fraction decomposition and fit the corresponding numerator ansatz.
  Numerator factors that have not been identified using the strategies
  of section~\ref{subsec:primdec} are parameterized via the ansatz
  of eq.~\eqref{eq:massive-ansatz}.
  Sampling on codimension one and two surfaces provides us with the
  numerical information needed to fix the free parameters in the ansatz.
\item Subtract the resultant expression and iterate until the whole coefficient has been reconstructed.
\end{enumerate}
  
\section{Amplitude for $\qb q Q \Qb H$}
\label{sec:amplitude}
Having given a succinct description of our analytic reconstruction techniques, we now turn to the application
to our specific process.
\subsection{Lowest order and definition of kinematics}
We study the reaction
\beq
0 \to \qb(p_1)+q(p_2)+Q(p_3)+\Qb(p_4)+H(p_5)\, ,
\eeq 
with $p_1+p_2+p_3+p_4+p_5=0$ and $p_1^2=p_2^2=0$, $p_3^2=p_4^2=m^2$, $p_5^2=m_H^2$.
The mass of the heavy quark, $Q$, is $m$. The quark $q$ is taken to be massless.
The colour structure of the process in lowest order is,
\beq
\label{LOqq}
\cA^{(0)}(1_{\qb},2_q,\three_Q,\four_{\Qb},\five_H) = i g^2 y_t (t^A)_{i_2i_1}  (t^A)_{i_3i_4} A^{(0)}(1_{\qb},2_q,\three_Q,\four_{\Qb},\five_H)  \, .
\eeq
In this equation the Yukawa coupling of the Higgs boson to the top quark is $y_t=m_t/v$ and $v$ is the Higgs vacuum
expectation value. The symbol $g$ denotes the strong coupling constant and $t^A$ are the $SU(3)$ colour matrices,
normalized so that ${\rm Tr}~t^A t^B=\delta^{AB}$.
Choosing a particular value for the helicities of the massless quarks the colour-stripped lowest order
amplitude has the simple form,
\beqn
A^{(0)}(1_{\bar{q}}^+,2_q^-,{\three}_Q,{\four}_{\bar{Q}},{\bf 5}_H)&=&
\frac{(\spa\three.\four+\spb\three.\four)}{s_{12}}\bigg[\frac{\spab2.\three.1}{\st_{123}}-\frac{\spab2.\four.1}{\st_{124}}\bigg] \nn\\
 &+&\bigg[\frac{\spb\three.1\spb1.\four}{\spb1.2}-\frac{\spa\three.2\spa2.\four}{\spa1.2}\bigg]
  \bigg[\frac{1}{\st_{123}}+\frac{1}{\st_{124}}\bigg]\, .
\eeqn
This is written using the spin-spinor notation of ref.~\cite{Arkani-Hamed:2017jhn} for the heavy quarks.
The $SU(2)$ little group index of the spin-spinors has been suppressed.
The amplitude for the opposite helicity combination of massless quarks is simply obtained by charge conjugation.
Squaring the lowest order amplitudes and summing over colours one obtains, 
\beq
\sum_{\rm colours} |\cA^{(0)}|^2 = g^4 y_t^2  V |A^{(0)}|^2\, .
\eeq
We define the $SU(N)$ colour charges as,
\beq
C_F=\frac{(N^2-1)}{2N}\,,\;\;\;V=(N^2-1)\,,\;\;\;N=3\,.
\eeq
\subsection{Amplitude at one-loop order}
In next-to-leading order we can decompose the colour structure as
follows, \beqn \label{oneloopcolourstructure}
\cA(1_{\qb},2_q,\three_Q,\four_{\Qb},\five_H) &=& g^4 y_t
\Big\{(t^A)_{i_2i_1} (t^A)_{i_3i_4}
A^{(1)}(1_{\qb},2_q,\three_Q,\four_{\Qb},\five_H) \nn \\ &+&
\delta_{i_2i_1} \delta_{i_3i_4}
A^{(1)}_{\delta\delta}(1_{\qb},2_q,\three_Q,\four_{\Qb},\five_H)\Big\}\, .
\eeqn
In evaluating the one-loop amplitude we include $n_{lf}$ light flavours
of (effectively) massless quarks in addition to the massive quark $Q$.
The interference between lowest order and next-to-leading order is,
\beq
\sum_{\rm colours} \;{\rm Re}~\{\cA^{(0)\,\dagger} \cA^{(1)}\}=
g^6 y_t^2  V \; {\rm Re}~\{A^{(0)\,\dagger} A^{(1)}\} \, .
\eeq
Note that $A^{(1)}_{\delta\delta}$ does not contribute to the matrix element squared at this order, so we will
drop it in the following.

The sub-amplitude $A^{(1)}$ can be expressed in terms of scalar integrals,
\begin{eqnarray} \label{fermionreduction}
{A}^{(1)}(1_{\qb}^{h_1},2_q^{h_2},\three_Q,\four_{\Qb};H) & = & \frac{\bar\mu^{4-n}}{r_\Gamma}\frac{1}{i \pi^{n/2}} \int {\rm d}^n \ell
 \, \frac{{\rm Num}(\ell)}{\prod_i d_i(\ell)} \nonumber \\
&= & \sum_{i,j,k} {d}_{i\x j\x k\x m_1 m_2 m_3 m_4}(1^{h_1},2^{h_2},\three,\four) \, D_0(p_i, p_j, p_k ;m_1,m_2,m_3,m_4)  \nonumber \\
&+& \sum_{i,j} {c}_{i\x j\x m_1 m_2 m_3}(1^{h_1},2^{h_2},\three,\four) \,  C_0(p_i,p_j ;m_1,m_2,m_3)   \nonumber \\
&+& \sum_{i} {b}_{i\x m_1 m_2}(1^{h_1},2^{h_2},\three,\four) \, B_0(p_i;m_1,m_2) + r(1^{h_1},2^{h_2},\three,\four)\, .
\end{eqnarray}

The scalar bubble ($B_0$), triangle ($C_0$) and box ($D_0$) integrals, together with the
constant $r_\Gamma$, are defined in Appendix~\ref{Integrals}.
$\bar{\mu}$ is an arbitrary mass scale, and $r$ are the rational terms.
The rank of a Feynman integral
is defined to be the number of powers of the loop momentum in the numerator. A scalar Feynman integral
has no powers of the loop momentum in the numerator, and is hence of rank zero.  All scalar
integrals are well known and readily evaluated using existing
libraries~\cite{Ellis:2007qk,vanHameren:2010cp,Carrazza:2016gav}.
Tadpole integrals are eliminated in favour of a scalar bubble integral and a rational term,
see Appendix~\ref{Integrals}, eq.~\eqref{eq:tadpole_elim}.
Scalar pentagon integrals can be reduced to the sum of 5 scalar box integrals,
see Appendix~\ref{pentagon_reduction}, eq.~\eqref{fivetofour}.

\subsection{Integral coefficient relations}

We define the following permutation operations:
\begin{eqnarray}
P_{2143}: && \left\{ 1 \leftrightarrow 2, \three \leftrightarrow \four,
  \spa{}.{} \leftrightarrow \spb{}.{} \right\} \nn \\
P_{2134}: && \left\{ 1 \leftrightarrow 2,
  \spa{}.{} \leftrightarrow \spb{}.{} \right\} \label{eq:permops} \\
P_{1243}: && \left\{\three \leftrightarrow \four \right\} \, . \nn
\end{eqnarray}
These operations can then be used to relate the coefficients of all necessary
integrals from the smaller basis set, as indicated in tables~\ref{tab:boxcoeff}--\ref{tab:bubcoeff}.
Coefficients in the second column are simply obtained by applying the operation $P_{2143}$.
Coefficients in columns 3 and 4 are obtained by applying the operations $P_{2134}$ and $P_{1243}$ and,
in addition, replacing an overall colour factor of $(N-2/N)$ by $2/N$.

\begin{table}
\centering
\begin{tabular}{l|l||l|l}
Integral coefficient & $P_{2143}$ & $P_{2134}$ & $P_{1243}$ \\
\hline
$d_{3 \x 4 \x 12 \x m0mm}$ & $d_{4 \x 3 \x 12 \x m0mm}$ & & \\
$d_{4 \x 2 \x 1 \x m000}$  & $d_{3 \x 1 \x 2 \x m000}$  
 & $d_{4 \x 1 \x 2 \x m000}$  & $d_{3 \x 2 \x 1 \x m000}$  \\
$d_{3 \x 1 \x 24 \x m00m}$ & $d_{4 \x 2 \x 13 \x m00m}$ 
 & $d_{3 \x 2 \x 14 \x m00m}$ & $d_{4 \x 1 \x 23 \x m00m}$  \\
$d_{3 \x 12 \x 4 \x m00m}$ & \\
\end{tabular}
\caption{
  The basic set of box integral coefficients which we calculate
  is listed in column 1. The coefficients in the other columns are obtained
  by the momentum permutations defined in eq.~\eqref{eq:permops}, and by modification of the colour factor (for columns 3 and 4).
\label{tab:boxcoeff}}
\end{table}
\begin{table}
\centering
\begin{tabular}{l|l||l|l}
Integral coefficient & $P_{2143}$ & $P_{2134}$ & $P_{1243}$ \\
\hline
$c_{1 \x 2 \x 000}$    & & & \\
$c_{3 \x 4 \x m0m}$    & & & \\
$c_{1 \x 3 \x 00m}$    & $c_{2 \x 4 \x 00m}$ & $c_{2 \x 3 \x 00m}$    & $c_{1 \x 4 \x 00m}$  \\
$c_{2 \x 13 \x 00m}$   & $c_{1 \x 24 \x 00m}$ & $c_{1 \x 23 \x 00m}$   & $c_{2 \x 14 \x 00m}$ \\
$c_{13 \x 24 \x m0m}$  & & $c_{14 \x 23 \x m0m}$ & \\
$c_{12 \x 3 \x 00m}$   & $c_{12 \x 4 \x 00m}$ & & \\
$c_{12 \x 3 \x mm0}$   & $c_{12 \x 4 \x mm0}$ & & \\
$c_{124 \x 3 \x m0m}$  & $c_{123 \x 4 \x m0m}$ & & \\
$c_{12 \x 34 \x mmm}$  & & & \\
\end{tabular}
\caption{
  The basic set of triangle integral coefficients which we calculate is listed in column 1.
  The coefficients in the other columns  are obtained
  by the momentum permutations defined in eq.~\eqref{eq:permops}
  and by modification of the colour factor (for columns 3 and 4).
\label{tab:tricoeff}}
\end{table}

\begin{table}
\centering
\begin{tabular}{l|l||l|l}
Integral coefficient & $P_{2143}$ & $P_{2134}$ & $P_{1243}$ \\
\hline
$b_{123 \x m0}$   & $b_{124 \x m0}$ & & \\
$b_{13 \x m0}$    & $b_{24 \x m0}$ & $b_{23 \x m0}$    & $b_{14 \x m0}$ \\
$b_{12 \x mm}$    & & & \\
$b_{1234 \x mm}$  & & & \\
$b_{12 \x 00}$    & & & \\
$b_{34 \x mm}$    & & & \\
$b_{3 \x m0}$     & & & \\
\end{tabular}
\caption{
The basic set of bubble integral coefficients which we calculate is listed in column 1.
Tadpole integrals are traded for $b_{3 \x m0}$ and rational terms. $b_{3 \x m0}$ itself
can be determined from the other bubble integral coefficients, because of the known
value of the ultraviolet poles of the one-loop amplitude. 
The coefficients in the other columns  are obtained
by the momentum permutations defined in eq.~\eqref{eq:permops}.
\label{tab:bubcoeff}}
\end{table}

\subsection{Choice of subtracted basis}
\label{subsec:subtractedbasis}

Although eq.~\eqref{fermionreduction} is a valid basis of scalar integrals in which to expand our amplitudes,
we find it convenient to use a basis in which the IR and collinear divergent box integrals have their singular
parts subtracted, which effectively moves all the IR pole divergences to the triangle sector. We denote the
subtracted box integrals by $\bar{D}_0$.  We note that they are closely related to 6-dimensional box integrals,
which would however also subtract contributions from infrared-finite triangles and would have a different overall factor.

There are three families of divergent box integrals which we must subtract.
The first is,
\beqn
&&\bar{D}_0(p_4,p_2,p_1,m,0,0,0) =D_0(p_4,p_2,p_1,m,0,0,0) 
-\frac{1}{\st_{24}} C_0(p_1,p_2,0,0,0) \nn \\
  &-&\frac{1}{s_{12}}C_0(p_2,p_4,0,0,m) +\frac{(s_{12}+\st_{14})}{s_{12} \st_{24}} C_0(p_1,p_{24},0,0,m)\, ,
\eeqn
together with the three other integral relations obtained by the permutations in eq.~\eqref{eq:permops}.
The second is,
\beqn
&&\bar{D}_0(p_3,p_1,p_{24},m,0,0,m)=D_0(p_3,p_1,p_{24},m,0,0,m)\nn \\
 &-&\frac{1}{\st_{124}} C_0(p_1,p_3,0,0,m) -\frac{(\st_{14}+s_{12})}{\st_{13} \st_{124}} C_0(p_1,p_{24},0,0,m)  \, ,
\eeqn
together with the three other integral relations obtained by the permutations in eq.~\eqref{eq:permops}.
The third is,
\beqn
&&\bar{D}_0(p_3,p_4,p_{12},m,0,m,m)= D_0(p_3,p_4,p_{12},m,0,m,m)\nn \\
&-&\frac{1}{\st_{124}}C_0(p_3,p_4,m,0,m)\, ,
\eeqn
together with the one other integral relation obtained by $\{3 \leftrightarrow 4\}$.

The coefficients of the box integrals are unchanged as a result of this subtraction, but after the subtraction of boxes,
the triangle coefficients become,
\beqn
\bar{c}_{1 \x 2 \x 000}  &=&    c_{1 \x 2 \x 000} + \frac{d_{4 \x 2 \x 1\x m000}}{\st_{24}}+ \frac{d_{4 \x 1 \x 2\x m000}}{\st_{14}}\nn\\
&+& \frac{d_{3 \x 2 \x 1\x m000}}{\st_{23}}+\frac{d_{3 \x 1 \x 2\x m000}}{\st_{13}},\\
\bar{c}_{1 \x 3 \x 00m} &=&    c_{1 \x 3 \x 00m} + \frac{d_{3 \x 1 \x 2\x m000}}{s_{12}}+\frac{d_{3 \x 1 \x 24 \x m00m}}{\st_{124}}, \nn\\ 
&+&\mbox{permutations}~P_{2143},P_{2134},P_{1243} \\
\bar{c}_{2 \x 13 \x 00m} &=&    c_{2 \x 13 \x 00m} - \frac{s_{12}+\st_{23}}{s_{12} \st_{13}} d_{3 \x 1 \x 2\x m000} + \frac{s_{12}+\st_{23}}{\st_{24} \st_{123}} d_{13 \x 2 \x 4\x m00m}, \; \nn \\
&+& \mbox{permutations}~P_{2143},P_{2134},P_{1243} \\
\bar{c}_{3 \x 4 \x m0m} &=&    c_{3 \x 4 \x m0m} +\frac{1}{\st_{124}} d_{3 \x 4 \x 12 \x m0mm}
+ \frac{1}{\st_{123}} d_{4 \x 3 \x 12 \x m0mm} \, .
\eeqn

\renewcommand{\baselinestretch}{1.4}
\begin{table}
\centering
\begin{tabular}{l|l||l}
Integral coefficient & Result & Related coefficients   \\
\hline
$\bar{c}_{1 \x 2 \x 000}$    & $ +\frac{1}{N} \; s_{12}       \; A^{(0)}(1_{\qb},2_q,\three_Q,\four_{\Qb},\five_H)$ &  \\
$\bar{c}_{1 \x 3 \x 00m}$    & $ -(N-\frac{2}{N}) \; \st_{13} \; A^{(0)}(1_{\qb},2_q,\three_Q,\four_{\Qb},\five_H)$ & $\bar{c}_{2 \x 4 \x 00m}$\\
$\bar{c}_{2 \x 3 \x 00m}$    & $ -\frac{2}{N} \; \st_{23}     \; A^{(0)}(1_{\qb},2_q,\three_Q,\four_{\Qb},\five_H)$   & $\bar{c}_{1 \x 4 \x 00m}$\\
$\bar{c}_{3 \x 4 \x m0m}$    & $ +\frac{1}{N} \; \st_{34}     \;A^{(0)}(1_{\qb},2_q,\three_Q,\four_{\Qb},\five_H)$& \\
$\bar{c}_{1 \x 23 \x 00m}$   & $0$ & $\bar{c}_{2 \x 14 \x 00m}$\\
$\bar{c}_{2 \x 13 \x 00m}$   & $0$ & $\bar{c}_{1 \x 24 \x 00m}$ \\
\end{tabular}
\caption{Subtracted triangle integral coefficients. All other triangle integral coefficients are unchanged.
  The basic set of coefficients is listed in the left-hand column, their value is in the middle column.
  The last column shows the coefficients obtained by the permutation operations,~eq.~\eqref{eq:permops}.
\label{tab:tricoeffbar}}
\end{table}
\renewcommand{\baselinestretch}{1}
The results for the subtracted triangle integral coefficients are given in Table~\ref{tab:tricoeffbar}.
The subtracted triangle coefficients are all proportional to the lowest order amplitude, or equal to zero.

\subsection{Ultraviolet renormalization}
The amplitudes presented so far are bare amplitudes, which require ultraviolet renormalization.
Our renormalization scheme has to satisfy three criteria,
\begin{itemize}
\item
The decoupling of heavy quarks should be manifest.
\item The evolution equations for the running coupling and for the 
parton distribution functions should be the same as the equations 
in the theory without the heavy quark. Both the strong coupling 
and the parton distribution functions should run with the 
coefficients appropriate for the $\overline{\rm MS}$ scheme in the absence of the massive particles.
\item The mass parameter should correspond to a pole mass.
\end{itemize}
These three criteria completely specify the renormalization scheme.
Full details of the renormalization scheme are given in ref.~\cite{Badger:2010mg}.

Introducing the renormalization of the bare parameters, $m_0 = Z_m m$ and $Q_0 = Z_Q Q$ we
may write the renormalized inverse propagator for a heavy quark of momentum $p$ as,
\beq
-i \Gamma_R(p,m;g)=Z_Q [ \slsh{p}-m-\Sigma(p,m;g)-m (Z_m-1)]+O(g^4) \, .
\eeq
$-i \Sigma(p,m;g)$ is the contribution of the one-loop heavy quark self-energy graph.
In the four-dimensional helicity (FDH) scheme we have,
\beqn
Z_Q &=& 1-g^2 \cg C_F \Bigg[\frac{3}{\e}+3 \ln\left(\frac{\mu^2}{m^2}\right)+5\Bigg]+O(g^4,\e)\, ,\\
Z_m &=& 1-g^2 \cg C_F \Bigg[\frac{3}{\e}+3 \ln\left(\frac{\mu^2}{m^2}\right)+5\Bigg]+O(g^4,\e)\, .
\eeqn
$Z_m$ is a purely ultra-violet effect, whereas the wave function renormalization of the heavy quark contains
both ultra-violet and infra-red divergences, both controlled by dimensional regularization~\cite{Badger:2010mg}.
Thus the agreement between the heavy quark mass renormalization and wave function renormalization should be considered
fortuitous.

The overall factor $\cg$ is,
\begin{equation}
\cg = \frac{1}{(4\pi)^{2-\e}}\frac{\Gamma(1+\e)\Gamma^2(1-\e)}{\Gamma(1-2\e)} = \frac{1}{16 \pi^2}\Big( 1 + (\ln(4 \pi) - \gamma_E) \e  + O(\e^2)\Big)\, ,
\end{equation}
where $\gamma_E$ is the Euler constant.
The coupling constant renormalization in the FDH scheme is,
\beq  \label{coupling_renormalization}
Z_g = 1- \frac{g^2}{16 \pi^2} \Big\{ \frac{1}{\epsilon} \Big[\frac{11}{3} N -\frac{2 n_{lf}}{3}\Big] - \frac{2}{3} \Big[\frac{1}{\epsilon}+\ln\Big(\frac{\mu^2}{m^2}\Big)\Big]\Big\}\,.
\eeq
where $n_{lf}$ is the number of massless quark flavours. The last term in eq.~\eqref{coupling_renormalization} gives
the contribution of the top quark.
Table~\ref{tab:counterterms} gives a full accounting of the one-loop counterterms which we have to add to our process.
\renewcommand{\baselinestretch}{1.7}
\begin{table}
\centering
\begin{tabular}{l|l}
Source of Counterterm &  Contribution \\
\hline
Mass renormalization                               & $ +g^2 \cg \Big( C_F \frac{3}{\epsilon}+3 \ln(\frac{\mu^2}{m^2})+5\Big) \, m \, A^{(ct)}$ \\
Yukawa coupling renormalization                    & $ -g^2 \cg \Big( C_F \frac{3}{\epsilon}+3 \ln(\frac{\mu^2}{m^2})+5\Big) \, A^{(0)}$ \\
Heavy quark wave function renormalization          & $ -g^2 \cg \Big( C_F \frac{3}{\epsilon}+3 \ln(\frac{\mu^2}{m^2})+5\Big) \, A^{(0)}$ \\
Coupling constant renormalization (light flavours) & $ +g^2 \cg \Big(-\frac{11}{3} N +\frac{4}{3} T_R n_{\lf}\Big)\frac{1}{\epsilon} \, A^{(0)}$ \\
Coupling constant renormalization (heavy flavour) & $ +g^2 \cg \frac{4}{3} T_R  \Big(\frac{1}{\epsilon}+\ln(\frac{\mu^2}{m^2})\Big) \, A^{(0)}$ \\
Translation to MS bar coupling                     & $ +g^2 \cg \frac{1}{3}N  \, A^{(0)}$ \\
\end{tabular}
\caption{Ultra-violet counterterms in the four dimensional helicity scheme.
\label{tab:counterterms}}
\end{table}
\renewcommand{\baselinestretch}{1}
The contribution to the amplitude of the mass counterterm is
\beqn
A^{(ct)}&=&
   \Big\{2 \* m \* \Big[\frac{(\spa\three.2 \* \spba1.\five.\four+\spb\three.1 \* \spab2.\five.\four)}{(\st_{123})^2}
         - \frac{(\spa2.\four \* \spab\three.\five.1+\spb1.\four \* \spba\three.\five.2)}{(\st_{124})^2}\Big] \nonumber \\
        &-&\big(\spa2.\four \* \spb\three.1+\spa\three.2 \* \spb1.\four\big)
            \* \Big[\frac{(s_{123}+3 \* m^2)}{(\st_{123})^2}+\frac{(s_{124}+3 \* m^2)}{(\st_{124})^2}\Big]\Big\}\frac{1}{s_{12}} \, .
\label{eq:Act}
\eeqn
\subsection{Infrared behaviour}
Further constraints on the amplitude can be obtained from the infra-red behaviour.
The one-loop singular behaviour of QCD amplitudes with massive partons was first addressed
by Catani et al, ref.~\cite{Catani:2000ef}. This reference also keeps track of certain finite terms, but we shall drop these in the following discussion. 
Subsequently, there have been great advances in the understanding of the infra-red behaviour of QCD amplitudes,
including in particular the extension to
higher orders~\cite{Sterman:2002qn,Becher:2009qa,Becher:2009kw,Ferroglia:2009ii,Ferroglia:2009ep}.
However since we are working at one-loop order, the results of ref.~\cite{Catani:2000ef}
will be sufficient for our purposes.
Ref.~\cite{Catani:2000ef} gives an expression for the pole structure of amplitudes in which
mass renormalization (and the associated Yukawa renormalization) has been performed, but
charge renormalization has not been performed.

We denote the mass-renormalized one-loop amplitude by ${\cal A}^{(1m)}$. 
Straightforward application of the formula in ref.~\cite{Catani:2000ef}
yields the following expression for the pole structure of the mass-renormalized amplitude,
\beqn
A^{(1m)}|_{\text{pole~part}}&=&\Bigg[-\frac{2C_F}{\e^2}+\frac{C_F}{\e} \left(2 \ln\left(\frac{-\st_{12}-i\varepsilon}{\mu^2}\right)-5\right) +\frac{\beta_0}{\e}
     +\frac{N}{ \e} \ln\left(\frac{|\st_{13}| |\st_{24}|}{m^2 (-s_{12}-i\varepsilon)}\right)\nn\\
  &+&\frac{1}{\e}\,\frac{1}{N}\left(\frac{1}{2 v_{34}}\ln\left(\frac{1-v_{34}}{1+v_{34}}\right)+\frac{i \pi}{v_{34}}\theta(\st_{34})
   -2 \ln\left(\frac{|\st_{13}| |\st_{24}|}{|\st_{23}| |\st_{14}|}\right) \right)
\Bigg] A^{(0)}\, ,
\eeqn
where,
\beq
v_{jk} = \sqrt {1 - \frac{4 m_j^2 m_k^2}{\st_{jk}^2}}\,,\;\;\;\beta_0=\frac{11}{3}N-\frac{2}{3} (n_{lf}+1)\,,
\eeq
and $\theta$ is the Heaviside theta function. This agrees with the result quoted in Ref.~\cite{Broggio:2015lya}.
This structure gives rise to the simple form of the subtracted triangle coefficients given in Table~\ref{tab:tricoeffbar}
and can be used to determine one of the bubble coefficients.

\section{Analytic forms of integral coefficients} 
All the integral coefficients have been computed using standard techniques, namely a combination of unitarity
cuts (see~\cite{Ellis:2011cr} and references therein)
and Passarino-Veltman reduction~\cite{Passarino:1978jh}, then
simplified using the methods of section~\ref{Reconstruction}.
The reconstruction can easily be checked at multiple phase-space points and for all spin choices to ensure
the correctness of the resultant expressions.  As a final check the full amplitude has been constructed, including
all renormalization factors, according to the procedure described in Section~\ref{sec:amplitude}.
Full agreement has been found when comparing with the automatic code Openloops~\cite{Buccioni:2019sur}.

In some cases the integral coefficients are sufficiently simple that they can be presented here in the text.
All coefficients in machine readable form can be extracted from the Fortran program which accompanies
the arXiv submission of this paper. They are also available in Python format on GitHub, Zenodo and the
PyPI, through the packages \textsc{Antares} and \textsc{Antares-Results} \cite{antares, antares_results}.
Human readable expressions for all coefficients are available in the
\href{https://gdelaurentis.github.io/antares-results/}{online documentation}. The evaluation at both a physical and at a finite-field phase space point is tested with \href{https://github.com/GDeLaurentis/antares-results/actions/workflows/ci_test.yml}{GitHub Actions}.

\subsection{Box integral coefficients}
Table~\ref{tab:boxcoeff} shows that there are 11 different
box-integral coefficients, of which 4 are independent. 
We give results for 3 out of the 4 independent box coefficients in the text.
\beqn
     && d_{3\x 4 \x 12\x m0mm}(1_{\qb}^+,2_q^-,\three_{Q},\four_{\Qb})=
\nn\\&&
-\frac{\spa\three.\four \* \spab2.\four.1 \* \st_{34}}{3 \* s_{12}}
-\frac{\spa2.\four \* \spa\three.2 \* \st_{34}}{3 \* \spa1.2}
\nn\\&&
+\frac{\st_{12\four} \* s_{34} \* (
 \spa\three.\four \* \spab2.\three.1 \* (\st_{34}-4 \* m^2)
 -3 \* m \* \spb1.\four \* (s_{34}-4 \* m^2) \* \spa\three.2)}{24 \* \Gramudxtxq}
\nn\\&&
+\frac{\st_{12\four}}{24 \* \spa1.2 \* \Gramudxtxq} \* \Big[
 (\spb\three.\four \* \spaba2.\three.\four.2 \* s_{34} \* (\st_{34}-4 \* m^2)
 - 2 \* m^3 \* \tr{1+2}.{3+4} \* (\spa2.\four \* \spba\three.\four.2
 + \spab2.\three.\four \* \spa\three.2))
\nn\\&&
+\spa\three.2 \* s_{34} \* m \* (\spaba2.\four.(1+2).\four \* m
 -\spa2.\four \* m \* (\st_{13}+\st_{23}+4 \* \st_{34})
 +\st_{13} \* \spab2.\three+\four.\four
 -(\st_{34}-4 \* m^2) \* \spab2.\three.\four)
\nn\\&&
+s_{34} \* (\spaba1.\three.\four.2 \* \spa2.\four \* \spab\three.\four.1
 -\spa\three.\four \* \spaba2.\three.\four.2 \* \st_{23}
 +6 \* (s_{34}-m^2) \* m^2 \* \spa2.\four \* \spa\three.2
\nn\\&&
 +(\st_{13}-\st_{34}-\st_{24}) \* \spa2.\four \* \st_{34} \* \spa\three.2)
-\spba\three.\four.2 \* \spa2.\four \* s_{34} \* m \* (
  3 \* \st_{34} - \spab2.\three+\four.2)
  \Big]
\nn\\&&
+\frac{{1}}{12 \* s_{12} \* \Gramudxtxq} \* \Big[s_{34} \* \st_{12\four} \* m \* (
 \spb1.\four \* \tr{1+2}.{3+4} \* \st_{34} \* \spa\three.2
 +\spb1.\four \* (\st_{34}-m^2) \* \st_{14} \* \spa\three.2
\nn\\&& \qquad\qquad
 +\spab\three.\four.1 \* (\spab1.\three.\four \* \spab2.\three+\four.1+\st_{23} \* \spab2.\three.\four)
 +\spb1.\four \* \st_{24} \* m \* \spba\three.\four.2-\spb\three.1 \* m \* \spab1.\three.\four \* \spab2.\four.1)\Big]
\nn\\&&
-\frac{\st_{12\four} \* \st_{34} \* (\spa2.\four \* \spab\three.\four.1 \* s_{34} \* (\st_{24}+\st_{14})
 -\st_{34} \* \spab2.\four.1 \* \spa\three.\four \* \tr{1+2}.{3+4})}{12 \* s_{12} \* \Gramudxtxq}
\nn\\&&
-\frac{m^3 \* \st_{12\four} \* \tr{1+2}.{3+4} \* (3 \* \spb1.\four \* \st_{34} \* \spa\three.2
 +2 \* \spab\three.\four.1 \* \spab2.\three.\four)}{12 \* s_{12} \* \Gramudxtxq}
\nn\\&&
-\frac{s_{34} }{96 \* s_{12} \* \Gramudxtxq^2} \* ((\spab2.\three.1 \* (\st_{14}-\st_{24})+\spab2.\four.1 \* (\st_{23}-\st_{13}))
 \* \st_{12\four}^2 \* (s_{34}-4 \* m^2) \nn\\&&
\qquad \times (m^2 \* \tr{1+2}.{3+4} \* (\spb\three.\four-\spa\three.\four)
  +m \* s_{34} \* (\spb\three.1 \* \spa1.\four+\spb\three.2 \* \spa2.\four-\spa\three.1 \* \spb1.\four-\spa\three.2 \* \spb2.\four)))
 \nn\\&&
 +\left\{ 1 \leftrightarrow 2, \spa{}.{} \leftrightarrow \spb{}.{} \right\}
 \, .
\eeqn
The box Gram determinant $\Gramudxtxq$ is as defined in eq.~\eqref{eq:boxgram} and
\beq
\tr{1+2}.{3+4} = \tilde s_{13} + \tilde s_{14} + \tilde s_{23} + \tilde s_{24} \, .\nn
\eeq

 \beqn
&&d_{4\x 2 \x 1\x m000}(1_{\qb}^+,2_q^-,\three_{Q},\four_{\Qb}) = \st_{24} \* (C_F-\frac{1}{2\*N}) \Bigg\{
\nn \\ 
 &+& \frac{(-2 \* (2 \* m \*\spb\three.1 +\spab{\three}.\five.{1})
 \* \spa{2}.\four-(2 \* m \* \spa\three.2 +\spba{\three}.{\five}.{2}) \* \spb{1}.\four)}{\st_{124}} \nn\\
&-&\frac{m \* \spa2.1 \* \spb{2}.\four \* \spab{\three}.\five.{1}
      +2 \* m \* \spa\three.1 \* \spb{2}.\four \* \spab{2}.\four.{1} + \st_{24} \* \spba{\three}.\five.{1} \* \spb1.\four }{\spab{1}.\four.{2} \* \st_{124}}\nn \\
&-&\frac{2 \* m^2 \* (\spa2.1 \* \spb{2}.\four \* \spb\three.1+\spa{2}.\four \* \spa\three.1 \* \spb2.1)}{\spab{1}.\four.{2} \* \st_{124}}
+ \frac{s_{12} \* \st_{24} \* \spb{2}.\four \* (-2 \* m \* \spa\three.1 -\spba{\three}.\five.{1})}{\spab{1}.\four.{2}^2 \* \st_{124}} \nn\\
&+&\frac{\st_{24} \* \spa1.2 \* \spb{2}.\four \* (2 \* m \* \spb1.2 \* \spa\three.1 
      -\spb\three.1 \* \spab{1}.\five.{2})}{\spab{1}.\four.{2} \* \spabab1.{\three}.(1+2).{\four}.2} 
+ \frac{\st_{24} \* ( \spb\three.1 \* \spab1.{\five}.4 + \spa\three.1 \* \spba1.{\five}.4)}{\spabab1.{\three}.(1+2).{\four}.2} \nn\\
&-&2 \* m \* \frac{ \st_{24} \* ( \spb\three.1 \* \spa1.\four + \spa\three.1 \* \spb1.\four)}{\spabab1.{\three}.(1+2).{\four}.2} \nn\\
&+&\frac{\spa{2}.\four \* \spa\three.2 \* \spbab1.{\three}.{\five}.1-\spb{1}.\four \* \spb\three.1 \* \spaba2.{\four}.{\five}.2
      + m \* (\spa\three.2 \* \spb{1}.\four+\spa{2}.\four \* \spb\three.1) \* (\spab{2}.\three.{1}-\spab{2}.\four.{1})}{\spabab2.{\four}.(1+2).{\three}.1}
      \Bigg\} \nn \\
\eeqn
 \beqn
&&d_{3\x 1 \x 24\x m00m}(1_{\qb}^+,2_q^-,\three_{Q},\four_{\Qb}) =\Big(C_F-\frac{1}{2N}\Big)\Bigg\{
\nn \\ 
 &-& \frac{2 \* \st_{124}^2 \* \spa\three.1 \* \spab{1}.\three.{\four} \* m \* \st_{13}}{\spaba1.{\five}.{\three}.1 \* \spbaba2.{\four}.(1+2).{\three}.1} \nn\\
&+&\Bigg[\spa\three.1 \* \Big(\spa{2}.\four \* \spbab1.{\three}.{\five}.1+2 \* m \* \spb{1}.\four \* \spab{2}.\three.{1}\Big) \nn \\
      &+&\spb\three.1 \* \spa{2}.\four \* m \* (\st_{15}+2 \* \st_{13}) +\spba\three.{\five}.2 \* \spb{1}.\four \* \st_{13} \Bigg] \nn\\
      &\times& \Bigg[\frac{\spbab1.{\three}.{\five}.1}{\spabab2.{\four}.(1+2).{\three}.1 \* \spb1.2}-\frac{\spaba1.{\five}.{\three}.1}{\spbaba2.{\four}.(1+2).{\three}.1 \* \spa1.2}\Bigg] \nn\\
&+&\st_{124} \* \spb\three.1 \* \Big(\frac{\spb{1}.\four \* \spab{2}.\five.{1} \* \st_{13}}{\spabab2.{\four}.(1+2).{\three}.1 \* \spb1.2}
      -\frac{\spab{1}.\three.{\four} \* \spaba2.{\five}.{\three}.1}{\spbaba2.{\four}.(1+2).{\three}.1 \* \spa1.2}\Big) \nn\\
&+&\frac{2 \* \spa1.2 \* \spb\three.1 \* \spb{1}.\four \* \spbab1.{\three}.{\five}.1 \* m^2}{\spabab2.{\four}.(1+2).{\three}.1 \* \spb1.2}\nn\\
&+&\frac{\st_{124} \* \spa\three.1 \* \Big(\spa{2}.\four \* m^2 \* (\st_{15}-2 \* \st_{13})
      +\spa{2}.\four \* \st_{13} \* s_{15}+m \* \spab{1}.\three.{\four} \* (2 \* \spab{2}.\three.{1}+\spab{2}.\five.{1})\Big)}{\spbaba2.{\four}.(1+2).{\three}.1 \* \spa1.2} \nn\\
&-&\frac{\spaba1.{\five}.{\three}.1 \* \st_{124} \* \spb\three.1 \* \spa{2}.\four \* m}{\spbaba2.{\four}.(1+2).{\three}.1 \* \spa1.2}\nn\\
&+&\frac{2 \* m \* (\st_{124} \* \spb\three.1 \* \spab{1}.\three.{\four} \* m-\spaba1.{\five}.{\three}.1 \* \spb\three.1 \* \spb{1}.\four \* m+\st_{124} \* \spa\three.1 \* \spb{1}.\four \* \st_{13})}{\spbaba2.{\four}.(1+2).{\three}.1}
\Bigg\}
\eeqn

 \subsection{Triangle integral coefficients}
Table~\ref{tab:tricoeff} shows that there are 19 different
triangle-integral coefficients, of which 9 are independent. 
After introducing the finite (triangle-subtracted) box integrals,
Table~\ref{tab:tricoeffbar} gives the result for 4 independent families of coefficients $\bar{c}$.
We present results for two of the remaining five independent coefficients here.
\beqn\label{eq:c12x34xmmm}
&&c_{12\x34\x mmm}(1_{\qb}^+,2_q^-,\three_{Q},\four_{\Qb})= \Bigg\{ \nn \\ &&
 \nn \\ &&
+\frac{s_{34}\*(\spabab2.\three.(1+2).\four.1-\spabab2.\four.(1+2).\three.1)\*m^2\*\Gramudxtq\*(3\*\spab\three.\four.2\*\spa2.\four
 -\spab\three.\four.1\*\spa1.\four-4\*\spa\three.\four\*\st_{23})}{12\*\spa1.2\*\spb1.2\*\Gramudxtxq^2}
 \nn \\ &&
+\frac{m^2\*\tr{1+2}.{3+4}\*(\spa\three.2\*\spa2.\four\*(3\*\st_{23}-\st_{13})+3\*\spa\three.1\*\spa2.\four\*\spab2.\four.1
 +2\*\spa\three.\four\*\spaba2.\four.(1+\three).2)}{2\*\spa1.2\*s_{34}\*\Gramudxtq}
 \nn \\ &&
-\frac{m\*\spa\three.2\*\tr{1+2}.{3+4}\*(\spb2.\four\*\spaba2.\three.\four.2+\spab2.\three.\four\*\st_{23}
 +\spab2.\three.\four\*\st_{13}+\spab1.\three.\four\*\spab2.\four.1)}{2\*\spa1.2\*s_{34}\*\Gramudxtq}
 \nn \\ &&
-\frac{m^2\*(\spa\three.1\*\spa2.\four\*\spab2.\four.1+2\*\spa\three.\four\*\spab2.\four.1\*\spa1.2
 +\spa\three.2\*\spa2.\four\*(3\*s_{12}+2\*\st_{23}+\st_{13}))}{\spa1.2\*\Gramudxtq}
 \nn \\ &&
+\frac{m\*\spa\three.2\*(\spb1.\four\*\spa1.2\*\spb1.2-\spbab1.\four.\three.\four
 -3\*\spb2.\four\*\spab2.\four.1-3\*\spb1.\four\*\st_{23})}{\Gramudxtq}
 \nn \\ &&
+\frac{m\*\spa\three.2\*\tr{1+2}.{3+4}\*(\spb2.\four\*\spaba2.\three.\four.2-2\*\spbab1.\four.\three.\four\*\spa1.2
 -3\*\spb1.\four\*\spa1.2\*m^2)}{12\*\spa1.2\*\Gramudxtxq}
 \nn \\ &&
+\frac{\tr{1+2}.{3+4}\*(2\*\spa2.\four\*\spaba1.\three.\four.2\*\spab\three.\four.1+2\*\spa\three.\four\*\st_{23}\*\spaba2.\three.\four.2
 -\spa\three.2\*s_{34}\*\spa2.\four\*(s_{34}-\four\*m^2))}{24\*\spa1.2\*\Gramudxtxq}
 \nn \\ &&
+\frac{\spb\three.\four\*\spaba2.\three.\four.2\*\tr{1+2}.{3+4}\*(s_{34}-\four\*m^2)}{24\*\spa1.2\*\Gramudxtxq}
 \nn \\ &&
+\frac{\spa\three.2\*s_{34}\*m\*(\spb1.\four\*m^2+4\*\spbab1.\four.\three.\four+\spb2.\four\*\spab2.\four.1
 +\spab2.\three.\four\*\spb1.2+\spb1.\four\*\st_{23}-3\*\spb1.\four\*\st_{34})}{6\*\Gramudxtxq}
 \nn \\ &&
+\frac{(\spabab2.\three.(1+2).\four.1-\spabab2.\four.(1+2).\three.1)\*s_{34}\*\spa\three.\four}{12\*\Gramudxtxq}
+\frac{\spab\three.\four.1\*s_{34}\*\spaba2.\four.\three.\four}{6\*\Gramudxtxq}
 \Bigg\} \nn \\ &&
+\left\{\three \leftrightarrow \four \right\} \nn \\ &&
+\left\{ 1 \leftrightarrow 2, \spa{}.{} \leftrightarrow \spb{}.{} \right\} \nn \\ &&
+\left\{ 1 \leftrightarrow 2, \three \leftrightarrow \four, \spa{}.{} \leftrightarrow \spb{}.{} \right\} \nn \\ &&
 \nn \\ &&
-\frac{\st_{34}\*s_{34}\*(\spabab2.\three.(1+2).\four.1-\spabab2.\four.(1+2).\three.1)
 \*\Gramudxtq\*m\*(\spab\three.(1+2).\four-\spba\three.(1+2).\four)}{12\*\spa1.2\*\spb1.2\*\Gramudxtxq^2}
 \nn \\ &&
+\frac{(\spabab2.\three.(1+2).\four.1-\spabab2.\four.(1+2).\three.1)\*\Gramudxtq\*\tr{1+2}.{3+4}\*m^\four
 \*(\spa\three.\four-\spb\three.\four)}{3\*\spa1.2\*\spb1.2\*\Gramudxtxq^2}
 \nn \\ &&
-\frac{3\*(\spab\three.(1+2).\four+\spba\three.(1+2).\four)\*\spab2.5.1\*\tr{1+2}.{3+4}\*s_{1234}\*m}{2\*\Gramudxtq^2}
\eeqn
This expression involves the corresponding triangle Gram determinant,
\beq
\Gramudxtq = \frac{\st_{13}+\st_{14}+\st_{23}+\st_{24}}{4} - s_{12}s_{34}
\eeq
 The second triangle coefficient is very simple,
\beqn\label{eq:c13x24m0m}
&&c_{13\x24\x m0m}(1_{\qb}^+,2_q^-,\three_{Q},\four_{\Qb})=
 (2C_F - \frac{1}{N}) \nn\\&& \times
 m \* \frac{\spab\three.\five.2\*\left(\spbab2.\five.(2+\four).\four
  -\spbab2.\four.\five.\four\right)-2\*\spab\three.(1+\three).\four \* \spbab2.\four.\five.2}{
 \spbab2.\four.\five.2 \* \spab1.\five.2} \nn\\&&
 +\left\{ 1 \leftrightarrow 2, \three \leftrightarrow \four,
  \spa{}.{} \leftrightarrow \spb{}.{} \right\} 
\eeqn
 \subsection{Bubble integral coefficients}
Table~\ref{tab:bubcoeff} shows that there are 11 different
bubble-integral coefficients, of which 6 are independent,
after taking into account that 
the bubble integral coefficient $b_{3 \x m0}$ can be
determined from the known form of the ultraviolet pole.
Specifically we have,
\beq \label{bubblesum}
b_{3 \x m0} = \left( C_F + \frac{11}{3}N-\frac{2}{3}(n_{lf}+1)\right) A^{(0)}
 - 3C_F m A^{(ct)} - \sum_{i \neq 3 \x m0} b_i\, ,
\eeq
where $A^{(ct)}$ has been defined in eq.~\eqref{eq:Act}.

We now present analytic results for 3 of the remaining 6 independent coefficients,
\beqn
&&b_{13\x m0}(1_{\qb}^+,2_q^-,\three_{Q},\four_{\Qb})=
 (2C_F - \frac{1}{N}) \bigg\{ \nn\\&&
\frac{m \* (\spa\three.2 \* \spb2.\four \* \st_{13} \* (\st_{13}+m^2)+\spa\three.1 \* \spb{1}.\four \* m^2 \* \spab{2}.{1+\three}.{2})}{
      \st_{123} \* \st_{13} \* \spab{1}.\three.{2} \* \spab{2}.{(1+\three)}.{2}}
\nn\\
      &+&\frac{((\spa2.\four \* \spb1.2+\spba1.\three.\four) \* m^2 \* \spa\three.1+\spa\three.\four \* (\st_{13}-m^2)^2)}{
      \st_{123} \* \st_{13} \* \spab{1}.\three.{2}}
\nn\\
      &-&\frac{((\st_{13}-m^2) \* (\spa\three.\four \* \spab{2}.\three.{1}+\spa\three.2 \* \spba1.\three.\four))}{
      \st_{123} \* \st_{13} \* \spab{2}.{(1+\three)}.{2}}
\nn\\
      &+&\frac{((2 \* \spa\three.2 \* \spba2.\three.\four-\spa\three.1 \* \spb1.2 \* \spa2.\four) \* \st_{13}^2
      -(\spa\three.\four \* \st_{23}+\spa\three.2 \* \spb1.2 \* \spa{1}.\four) \* m^4)}{
      \st_{123} \* \st_{13} \* \spab{1}.\three.{2} \* \spab{2}.{(1+\three)}.{2}}
\nn\\
      &-&\frac{m \* (m \* \spb\three.\four+\spba\three.(1+2).\four)}{\st_{123} \* \spab{1}.\three.{2}}
\bigg\}
\eeqn
 \beqn \label{b12xmm}
b_{12\x mm}(1_{\qb}^+,2_q^-,\three_{Q},\four_{\Qb})&=&
 \Big\{\frac{-s_{12} \* m \* \spab{2}.\three.{1} \* (\st_{(1+2)3}+4 \* m^2)
       \* (2 \* \spb{2}.\four \* \spa\three.2+\spb{1}.\four \* \spa\three.1)}{32 \* \Gramudxt^2 \* \st_{123})}\nn\\
&+&\frac{s_{12} \* \spab{2}.\three.{1} \* (\st_{(1+2)3}+4 \* m^2)
       \* (\spba2.\three.\four \* \spa\three.2+\st_{13} \* \spa\three.\four)}{32 \* \Gramudxt^2 \* \st_{123}}\nn\\
&+&\frac{3 \* \spa\three.2 \* \spb{2}.\four \* \spab{2}.\five.{1} \* m \* (2 \* s_{12}+2 p_{12}.p_{34})}{2 \* \Gramudxtq^2}\nn\\
&+&\frac{4 \* m^2 \* (-\spa\three.2 \* \spa{2}.\four \* \spb1.2-\spab{2}.\three.{1} \* \spa\three.\four)}{3 \* s_{12}^2 \* \st_{123}}\nn\\
&+&\frac{(\spb1.2 \* m \* \spa\three.2 \* (\spab{2}.\three.{\four}-\spa{2}.\four \* m))}{s_{34} \* \Gramudxtq}\nn\\
&+&\frac{\spb1.2 \* \spa\three.2 \* m \* (4 \* \spab{2}.\three.{\four}-3 \* \spa1.2 \* \spb{1}.\four)}{24 \* \Gramudxt \* \st_{123}}\nn\\
&+&\frac{(4 \* \spa\three.\four \* \spab{2}.\three.{1} \* (\st_{(1+2)3}+2 \* m^2)
      +\spb1.2 \* \spa{2}.\four \* \spaba\three.(1+2).\three.2)}{24 \* \Gramudxt \* \st_{123}}\nn\\
&+&\frac{(\spa\three.2 \* \spb{1}.\four \* m)}{\Gramudxtq}\nn\\
&+& G(1,2,\three,\four)
\Big\} \nn \\
&+&\left\{\three \leftrightarrow \four \right\}
+\left\{ 1 \leftrightarrow 2, \spa{}.{} \leftrightarrow \spb{}.{} \right\}
+\left\{ 1 \leftrightarrow 2, \three \leftrightarrow \four, \spa{}.{} \leftrightarrow \spb{}.{} \right\}
\eeqn
Here we have defined an auxiliary function containing all of the 
$\Gramudxtxq$ poles,
\beqn \label{b12xmmGbit}
G(1,2,\three,\four) &=&
\Big[( m^2 \* \spa\three.2 \* \spa{2}.\four \* \spb1.2
       + 2 \* \spa\three.\four \* \spab{2}.\four.{1} \* (\st_{13} - \st_{23}) \nn\\
      &-&\spab{2}.\three.{1} \* (\spab{3}.\four.{1} \* \spa{1}.\four-4 \* \spa\three.\four \* \st_{34}
      -3 \* \spa\three.\four \* (\st_{24} - 2 \* m^2)))\nn\\
&-&  m \* (6 \* \spab{2}.\three.{\four} \* \spab{3}.\four.{1}+4 \* \spab\three.{(1+2)}.\four \* \spab{2}.\four.{1}
                    - \spa\three.2 \* \spb{1}.\four \* \st_{13} \nn\\
      &+&8 \* \spa\three.2 \* \spb{1}.\four \* (\st_{34}-m^2))\Big]
      \times \frac{F_2(p_1,p_2,p_3,p_4)}{96 \* \Gramudxt \* \Gramudxtxq}\nn\\
&-&\frac{\st_{34} \* \spa\three.\four \* \spab{2}.\three.{1} \* F_1}{96 \* \Gramudxt \* \Gramudxtxq}\nn\\
&+&\frac{((\spab{2}.\three.{\four} \* \st_{13} \* \spa\three.2
      -\spa{2}.\four \* \spab{2}.\three.{1} \* \spba{3}.\four.{1}) \* m \* F_2(p_1,p_2,p_3,p_4))}{48 \* \spa1.2 \* \Gramudxt \* \Gramudxtxq}\nn\\
&+&\frac{m \* \spaba2.\three.\four.2 \* \spba\three.{(1+2)}.\four \* F_1 }{24 \* \spa1.2 \* \Gramudxt \* \Gramudxtxq}\nn\\
&+&\frac{ F_2(p_1,p_2,p_3,p_4)
       \* (\spa{2}.\four \* \st_{34} \* \st_{13} \* \spa\three.2-(\spb\three.\four \* \st_{(1+2)3}
       +\spa\three.\four \* \st_{23}) \* \spaba2.\three.\four.2)}{48 \* \spa1.2 \* \Gramudxt \* \Gramudxtxq}\nn\\
&+&\frac{\spa\three.2 \* m \* (m \* \spa1.2 \* \spba{1}.\three.\four+2 \*m^2 \* \spa1.2 \*\spb1.\four 
      -2 \* \spaba2.\three.\four.2 \* \spb{2}.\four)}{12 \* \spa1.2 \* \Gramudxtxq}\nn\\
&-&\frac{(\spa{2}.\four \* \spab{3}.\four.{1} \* \spaba1.\three.\four.2+(\spab{3}.\four.{1}
        \* \spa{1}.\four+2 \* \spa\three.\four \* \st_{23}) \* \spaba2.\three.\four.2}{12 \* \spa1.2 \* \Gramudxtxq}
\eeqn
and also,
\beqn 
F_1&=&\st_{(1+2)3}\* \st_{(1+2)4}-2 \* s_{12} \* \st_{34} \label{b12xmmFbit1} \, ,\\
F_2(p_1,p_2,p_3,p_4)&=& \st_{34} \st_{(1+2)3} \*-2\* m^2 \* \st_{(1+2)4} \, , \label{b12xmmFbit2}
\eeqn
with $\st_{(1+2)j}=2 p_{1}\cdot p_j+2 p_{2}\cdot p_j$.
 
The function $G(1,2,\three,\four)$ also appears in the coefficient $b_{12\x 00}$,
\beqn \label{b12x00}
&&b_{12\x 00}(1_{\qb}^+,2_q^-,\three_{Q},\four_{\Qb})= \nn\\
&& \Big\{
\frac{9\*\spab2.\three.1\*s_{12}\*(m^2\*s_{12}\*\spa\three.\four-m\*\spabab\three.{(1+2)}.\three.{(1+2)}.\four)}{16\*\Gramudxt^2\*\st_{12\three}}
 \nn \\ &&
 - 9 \, G(1,2,\three,\four)
 	\Big\}
 \nn \\
&+&\left\{\three \leftrightarrow \four \right\}
+\left\{ 1 \leftrightarrow 2, \spa{}.{} \leftrightarrow \spb{}.{} \right\}
+\left\{ 1 \leftrightarrow 2, \three \leftrightarrow \four, \spa{}.{} \leftrightarrow \spb{}.{} \right\}
\nonumber \\
\nonumber \\
&+& \Big\{
       \frac{{1}}{24\*\Gramudxt\*\st_{12\three}} \* \Big(    
        -4\*m^2\*\spb\three.1\*\spab2.\three.\four+48\*\spa\three.2\*m\*\st_{23}\*\spb1.\four
       +12\*\spa\three.1\*m\*\spab2.\three.1\*\spb1.\four
 \nonumber \\ && \qquad
       -8\*\spa1.2\*\spb\three.1\*(\st_{13}\*\spb1.\four+\spb1.2\*\spa2.\four\*m)
       +76\*m^2\*\spa\three.\four\*\spab2.\three.1
 \nonumber \\ && \qquad
       +20\*\spb1.2\*\spa\three.2\*m^2\*\spa2.\four
       -28\*\spb1.2\*\spa\three.1\*\spa2.\four\*\spab2.\three.1-\four9\*\spb1.2\*\spa1.2\*\spa\three.2\*m\*\spb1.\four \Big)
 \nonumber \\      
       &-&\frac{(17\*\spab2.\three.1\*\spb\three.\four+\spa\three.2\*m\*\spb1.\four
       +37\*\spa\three.\four\*\spab2.\three.1)\*(\st_{13}+\st_{23})}{24\*\Gramudxt\*s_{12}}
 \nonumber \\    
       &+&\frac{{1}}{24\*\Gramudxt}\* \Big( -\frac{21\*\spb\three.1\*\spb1.\four\*(\st_{13}+\st_{23})}{\spb1.2}
       +\frac{23\*\spa2.\four\*\spa\three.2\*(\st_{13}+\st_{23})}{\spa1.2}
 \nonumber \\ && \qquad \qquad \qquad
       -\spab2.\three.1\*\spb\three.\four-65\*\spa\three.2\*m\*\spb1.\four-36\*\spa2.\four\*\spb\three.1\*m
       +19\*\spa\three.\four\*\spab2.\three.1 \Big)
 \nonumber \\       
       &+&\frac{43\*\spa\three.\four\*\spab2.\three.1+\spa\three.2\*(24\*\spb1.\four\*m+29\*\spb1.2\*\spa2.\four)
        +\spb\three.1\*(23\*\spab2.\three.\four-27\*\spa1.2\*\spb1.\four)}{6\*s_{12}\*\st_{12\three}}
 \nonumber \\ 
       &+&\frac{(\st_{13}-\st_{23})}{12\*\spab1.\three.2\*\Gramudxt\*\st_{12\three}} \* \Big(
        9\*\spb2.\four\*\spa\three.1\*m\*\spab2.\three.1-9\*\spa1.2\*\spb\three.1\*m\*\spba2.\three.\four
 \nonumber \\ && \qquad
       -2\*\spa1.2\*\spb2.\four\*\spb\three.2\*\spab2.\three.1
       -7\*\spb1.2\*\spa\three.1\*\spa1.\four\*\spab2.\three.1
       -9\*\spb1.2\*m^2\*\spa1.\four\*\spa\three.2 \Big)
\Big\}
 \nn \\
&+&\left\{ 1 \leftrightarrow 2, \three \leftrightarrow \four, \spa{}.{} \leftrightarrow \spb{}.{} \right\}
\nn \\
&-&\frac{2n_{lf}}{3} A^{(0)}
\eeqn
Eqs.~(\ref{b12xmm}) and (\ref{b12x00}) have demonstrated a close relationship between the bubble coefficients, in particular
for the terms involving inverse powers of the Gram determinant, $\Gramudxtxq$. Because of \eqref{bubblesum} it is clear
that all such terms must cancel in the sum of the bubble coefficients. It is therefore likely that further simplifications
can be obtained by identifying these terms in the remaining bubble coefficients.
\section{Conclusions}
We have performed the first application of analytic reconstruction to
amplitudes with massive quarks using the spin-spinor formalism.
This method has
interest for one-loop calculations and also potentially for two-loop
amplitudes involving heavy quarks, where simplification and reconstruction methods are
much more crucial.

It is still possible that even simpler results could be obtained
with more sophisticated forms for the factorized polynomials in the numerators of the integral coefficients.
Our Fortran code for the full amplitude is faster, by about a factor of two
than the result obtained, for the same process, with
the automatic code Openloops~\cite{Buccioni:2019sur}.  However this falls short of
the improvements achieved with other simplified amplitudes, such as the factor
of an order of magnitude or more achieved in refs.~\cite{Budge:2020oyl,Campbell:2022qpq,Campbell:2021vlt}. This is
a symptom of the complexity still remaining in our expressions.
It remains to be seen whether this complexity is intrinsic to a five-point
amplitude with 3 massive external particles, or whether further simplifications
can be achieved. Machine readable expressions for all the integral coefficients
can be extracted from the Fortran program which accompanies this paper, or from
the Python implementation available on GitHub~\cite{antares_results}.

The extension of this method to a complete physical process at one-loop order
would require calculation of the $ggt\bar{t}H$ amplitude. This is not expected
to present new conceptual problems, although the presence of both colour-leading
and colour-suppressed amplitudes will increase the number of coefficients that needs
to be calculated.

\acknowledgments

We thank Rafael Aoude and Vasily Sotnikov for useful discussions.
RKE acknowledges receipt of a Leverhulme Emeritus Fellowship from the
Leverhulme Trust.  RKE also wishes to thank the CERN Theoretical
Physics Department and the Fermilab Theoretical Division for
hospitality while part of this work was carried out.
GDL's work is supported in part by the U.K.\ Royal Society through
Grant URF\textbackslash R1\textbackslash 20109.
The work of JMC is supported in part by the U.S. Department of Energy, Office of Science, Office of Advanced Scientific Computing Research, Scientific
Discovery through Advanced Computing (SciDAC-5) program, grant ``NeuCol''.
This manuscript has been authored by FermiForward Discovery Group, LLC under
Contract No. 89243024CSC000002 with the U.S. Department of Energy,
Office of Science, Office of High Energy Physics.

\appendix

\section{Spinor techniques}
\label{Spinor_techniques}
\subsection{Massless spinors}
We work in the metric given by ${\rm diag}(1,-1,-1,-1)$ and use the Weyl representation of the Dirac
gamma matrices given by\footnote{Other authors may use a different definition of the Weyl representation
where the fields with left handed chirality occupy the
upper two components of the Dirac spinors.}
\beq \label{gammamatrices}
\gamma^\mu = \left(\begin{matrix}{\bf 0}&{\bar{\boldsigma}}^\mu\cr
  {\boldsigma}^\mu&{\bf 0}\cr \end{matrix}\right)\ ,(\mu=0,3),\,\,
\gamma_5 = i \gamma^0\gamma^1\gamma^2\gamma^3 =
\left(\begin{matrix}{\bf 1}&{\bf 0}\cr{\bf 0}&-{\bf 1}\cr \end{matrix}\right)\ ,
\eeq
where $\boldsigma^\mu=({\bf 1},{\boldsigma}^i)$, ${\bar{\boldsigma}}^\mu=({\bf 1},-{\boldsigma}^i)$
where $\boldsigma$ are the Pauli matrices,
\beq
{\boldsigma}^1 =\left(\begin{matrix} 0 & 1 \\ 1 & 0 \end{matrix}\right),\;\;
{\boldsigma}^2 =\left(\begin{matrix} 0 & -i \\ i & 0 \end{matrix}\right),\;\;
{\boldsigma}^3 =\left(\begin{matrix} 1 & 0 \\ 0 & -1 \end{matrix}\right)\, .
\eeq
In the representation of \eqref{gammamatrices}
we have
\beq
\slsh{p} =p^0 \gamma^0-p^1 \gamma^1-p^2 \gamma^2-p^3 \gamma^3 = 
\left(\begin{matrix}
  0& 0 &  p^+ & \bar{p}_\perp \\
  0& 0 &  p_\perp & p^- \\
  p^- & -\bar{p}_\perp & 0 & 0 \\
  -p_\perp & p^+ & 0 & 0  \end{matrix}\right) = \left(\begin{matrix} {\bf 0} & p_{ \alpha \db} \\
                                                                     p^{ \da \beta} & \bf{0} \end{matrix}\right)\, ,
\eeq 
with
\beq
\label{masslesscase}
p^\pm=p^0\pm p^3, \,p_\perp=p^1+i p^2, \,\bar{p}_\perp=p^1-i p^2.
\eeq
Note that the determinants of $ p_{ \alpha \db}$ and  $p^{ \da \beta}$ are equal to zero.
In the calculation of amplitudes we take all particles to be outgoing.
Therefore the ingredients that we require are the plane wave solutions of the Dirac equation for outgoing
fermions, $\bar{u}(p)$ and outgoing anti-fermions, $v(p)$.
For light-like vectors the Dirac spinors in the Weyl representation 
can be decomposed into two-dimensional Weyl spinors
as follows,
\beqn
\bar{u}_{+}(p)&=&
\left(\begin{matrix} \langle 0|^\alpha,& [p|_\da \end{matrix}\right)\,
=\left(\begin{matrix} 0,& 0, \sqrt{p^+},& \bar{p}_{\perp}/\sqrt{p^+} \end{matrix}\right)=\bar{v}_{-}(p)\, ,\nn \\
\bar{u}_{-}(p)&=&\left(\begin{matrix} \langle p|^\alpha,& [0|_\da\end{matrix}\right)
=\left(\begin{matrix} {p}_{\perp}/\sqrt{p^+},& -\sqrt{p^+},& 0,& 0\end{matrix}\right)=\bar{v}_{+}(p)\, ,
\eeqn
\beq
v_{+}(p)=\left(\begin{matrix} |0\rangle_{\beta}, \\
                              |p]^{\db}, \end{matrix}\right)
=\left(\begin{matrix} 0\\
                      0\\
                      \bar{p}_{\perp}/\sqrt{p^+}\\
                      -\sqrt{p^+}\end{matrix}\right)\, = u_{-}(p)\,,\quad
v_{-}(p)=\left(\begin{matrix} |p\rangle_{\beta}\\
                              |0]^{\db} \end{matrix}\right)
=\left(\begin{matrix}    \sqrt{p^+}\\
                         p_{\perp}/\sqrt{p^+}\\
                         0 \\
                         0                  \end{matrix}\right) =u_{+}(p)\, .
\eeq
The angle (square) brackets automatically encode the north-west~$\to$~south-east
(south-west~$\to$~north-east) summation convention 
for the SL$(2,\mathbb{C})$ dotted (undotted) indices.
Thus in most circumstances these indices can be dropped.
For massless particles the $p_\mu p^\mu={\rm det}~p_{\alpha \db}=0$ and the matrices
can be expressed as bi-spinors,
\beq \label{bispinors}
p^{\da \beta}=|p]^\da \langle p|^{\beta},\;\;
p_{\alpha \db}=|p\rangle_\alpha [ p|_{\db}\, .
  \eeq
  The $u$ and $v$ spinors satisfy the charge conjugation relations,
$v_{\pm}(p)=C \bar{u}_{\pm}^T$ and $u_{\pm}(p)=C \bar{v}_{\pm}^T$,
\beqn
{ C} &=& -C^{-1}=
{ -i \gamma^2 \gamma^0 = \ \left(\begin{matrix} i{\boldsigma}^2& {\bf0}
\cr                                      {\bf0}& -i{\boldsigma}^2\cr \end{matrix}\right)}
=   \left(\begin{matrix}
      0& 1 & & 0& 0 \cr
      -1& 0 & & 0& 0 \cr
      0& 0 & & 0& -1 \cr
      0& 0 & & 1& 0 \cr \end{matrix}\right) \,
=\left(\begin{matrix} \epsilon^{\da \db}
  & {\bf0}\\
  {\bf 0} & \epsilon_{\alpha\beta} \end{matrix}\right)\, ,
\eeqn
where the two-dimensional antisymmetric tensor is,
\beq 
\epsilon^{\alpha \beta} = \epsilon^{\da \db} = i \boldsigma^2 =\left(\begin{matrix} 0 & 1 \\
                                              -1 & 0 \end{matrix}\right),\;\;\;
\epsilon_{\alpha \beta}= \epsilon_{\da \db}= -i \boldsigma^2 =\left(\begin{matrix} 0 & -1 \\
                                               1 & 0 \end{matrix}\right) \, .
\label{eq:epdef}
\eeq
Thus to raise or lower the index of a spinor quantity, adjacent spinor
indices are summed over when multiplied on the left by the appropriate epsilon symbol,
\beq
|p\rangle_{\alpha}=\epsilon_{\alpha \beta}\langle p|^{\beta}, \quad
|p]^\da=\epsilon^{\da \db}[p|_{\db}\, ,
\eeq
and analogously,
\beq
\langle p|^{\alpha}=\epsilon^{\alpha \beta}|p\rangle_{\beta}, \quad
[p|_\da=\epsilon_{\da \db}|p]^{\db}\, ,
\eeq
and,
\beq
p_{\alpha \db}=\epsilon_{\alpha \beta} \epsilon_{\db \da}p^{\da \beta} \,.
\eeq
Using eq.~\eqref{bispinors} we see that the massless spinors satisfy the Weyl equations of motion,
\beq
p^{\da \beta} |p\rangle_\beta = p_{\alpha \db} |p]^\db =
\langle p|^{\alpha} p_{\alpha \db}=[ p|_{\da} p^{\da \beta}=0 \,.
\eeq
For real momenta we have that,
\beq
(\langle p |^\alpha )^* = | p]^\da\,,\;\;\;\;\; ([p|_\da)^*= |p \rangle_\alpha \, .
\eeq
\subsection{Massive spinors}
For the case of massive spinors the determinant of $p$ is no longer equal to zero.
But following ref.~\cite{Arkani-Hamed:2017jhn} we can represent $p$ as the sum of the bispinor products
of two massless momenta\footnote{The representation of a massive spinor in terms of two massless spinors
has first been suggested in ref.~\cite{Kleiss:1985yh}, where a particular choice of the massless spinors is made.
Ref.~\cite{Arkani-Hamed:2017jhn} argues that making this choice at an early stage obscures
the little group properties of the amplitude.}.
\beq
p^{\da \beta}= \sum_{I=1,2} |\bm{p}_I]^{\da} \langle \bm{p}^I |^\beta\, , \;\;\;\;\;
p_{\alpha \db}= \sum_{I=1,2} |\bm{p}_I]_{\alpha} \langle \bm{p}^I |_\db\; .
\eeq

In our notation $I$ and $J$ taken on the values $1$ and $2$.
Raising and lowering the $SU(2)$ index is performed
by multiplying by the two-dimensional totally antisymmetric tensor $\epsilon$ on the right,
$|\bm{p}_I\rangle_{\alpha} = |\bm{p}^J\rangle_{\alpha}\epsilon_{JI}$
and $\lbrack\bm{p}^I|_{\da} = \lbrack\bm{p}_J|_{\da}\epsilon^{JI}$.
For real momenta we have,
\beqn
\Big( \langle \bm{p}^I|^\alpha \Big)^* = |\bm{p}_I]^\da \,, \qquad
\Big( | \bm{p}^I \rangle_\alpha \Big)^* = [\bm{p}_I|_\da \,,
\eeqn
but on the other hand,
\beqn
\Big( \langle \bm{p}_I|^\alpha \Big)^* = - |\bm{p}^I]^\da \,, \qquad
\Big( | \bm{p}_I \rangle_\alpha \Big)^* = - [\bm{p}^I|_\da \,.
\eeqn
In other words, taking the complex conjugate of an angle spinor with a lowered spin index $I$, or a square
spinor with a raised spin index, introduces an additional minus sign.
This means that if we define the spinors for an outgoing quark and antiquark as,
\beq
\bar{u}(p) = \left(
\begin{matrix} \langle \bm{p}^I |^\alpha & \; {[}\bm{p}^I |_\da 
\end{matrix}
\right) \,, \qquad
v(p) = \left(
\begin{array}{c}  |\bm{p}_I \rangle_\alpha \\ | \bm{p}_I ]^\da \\ \end{array}
\right) \,,
\label{eq:ubarandv}
\eeq
then we must also have,
\beq
u(p) = \left(
\begin{array}{c} - | \bm{p}_I \rangle_\alpha \\ | \bm{p}_I ]^\da 
\end{array}
\right) \,, \qquad
\bar v(p) = \left(
\begin{matrix} \; \langle \bm{p}^I |^\alpha & -[ \bm{p}^I |_\da \end{matrix}
\right) \,.
\eeq

\section{Massive Spinors in \textsc{lips} and \textsc{antares}}
In this appendix, we give details of the Python code \textsc{lips}
\cite{DeLaurentis:2023qhd, Lips}, which allows the generation of
phase-space configurations defined either in terms of four-momenta or
of Weyl spinors.  The package generates these configurations, not only
over complex numbers ($\mathbb{C}$), but now also over finite fields
($\mathbb{F}_p$) and $p\kern0.2mm$-adic numbers ($\mathbb{Q}_p$).
More technical details are available in
ref.~\cite{DeLaurentis:2023qhd}.  The code uses
\textsc{Singular}~\cite{DGPS}, a computer algebra system for
polynomial computations in algebraic geometry, through the Python
interface \textsc{Syngular} \cite{syngular}. Number types for finite
fields and $p\kern0.2mm$-adic numbers are implemented in
\textsc{Pyadic} \cite{pyadic}.  We also demonstrate the usage of
\textsc{Antares} \cite{antares} to load and evaluate the integral
coefficients that are the main subject of this work. The required
language is Python 3, with version $\geq 3.9$ being recommended.
\label{lips}

The packages can be installed from the Python Package Index from the terminal:
\begin{minted}[escapeinside=||, mathescape=true, linenos=False, numbersep=5pt, gobble=2, frame=lines, framesep=2mm, breaklines, breakautoindent=false, breakindent=-12.5pt]{bash}
pip install lips antares
\end{minted}
This will install \textsc{Lips}, \textsc{Antares} and their
dependencies, such as \textsc{Syngular} and \textsc{Pyadic}.  The
number field, \texttt{Field}, is defined in \textsc{Syngular}, while
the phase space object, \texttt{Particles}, is defined in
\textsc{Lips}. The following code snippets are provided also as an
ancillary file (\texttt{lips-massive-spinors-example.ipynb}), which is
a Jupyter notebook that demonstrates these operations in practice.
We first import these objects:
\begin{minted}[escapeinside=||, mathescape=true, linenos=False, numbersep=5pt, gobble=2, frame=lines, framesep=2mm, breaklines, breakautoindent=false, breakindent=-12.5pt]{python}
from syngular import Field
from lips import Particles
\end{minted}

The code snippet given below shows the generation of an eight-point massless phase
space point as complex numbers (field characteristic 0, and we choose 300
digits of precision) and subsequent clustering into the five-point
massive phase-space. A completely generic momentum-conserving
eight-point phase space is generated, and then the mass constraints of
eqs.~\eqref{eq:mass-constraints-first}-\eqref{eq:mass-constraints-last}
are applied via the \texttt{\_singular\_variety} function. The
constraints should be satisfied to the full working precision of 300
digits. In the \texttt{cluster} function, we specify the map described
by eq.~\eqref{eq:eight-point-to-five-point-first}, as well as which
leg corresponds to a massive spinor after the clustering, the position
of its $SU(2)$ index (\texttt{u} for up, \texttt{d} for down) and its
value (either 1 or 2):
\begin{minted}[escapeinside=||, mathescape=true, linenos=False, numbersep=5pt, gobble=2, frame=lines, framesep=2mm, breaklines, breakautoindent=false, breakindent=-12.5pt]{python}
# 300-digits complex floating-point phase space point
oPs8ptC = Particles(8, field=Field("mpc", 0, 300), seed=0)
oPs8ptC._singular_variety(("⟨34⟩+[34]", "⟨34⟩-⟨56⟩", "⟨56⟩+[56]"),
                          (10 ** -300, 10 ** -300, 10 ** -300), seed=0)
oPs8ptC.mt = oPs8ptC.m_t = - oPs8ptC("⟨34⟩")
oPs8ptC.mh = oPs8ptC.m_h = oPs8ptC.field.sqrt(oPs8ptC("s_78"))
oPs5ptC = oPs8ptC.cluster([[1, ], [2, ], [3, 4], [5, 6], [7, 8]],
                          massive_fermions=((3, 'u', 1), (4, 'd', 1)))
\end{minted}
We demonstrate the subsequent evaluation of an expression defined in
five-point massive kinematics with 300 digit accuracy. The spinor
notation should be self-explanatory and $\Delta_{12|3|4|5}$ and
$\Delta_{12|34|5}$ are the box and triangle Gram determinants as
defined in eqs.~\eqref{eq:boxgram} and \eqref{eq:trigram}
respectively.  For ease of use the massive spinors $\three$ and
$\four$ are displayed without bold face, but this should be understood
(and is also supported through unicode characters):
\begin{minted}[escapeinside=||, mathescape=true, linenos=False, numbersep=5pt, gobble=2, frame=lines, framesep=2mm, breaklines, breakautoindent=false, breakindent=-12.5pt]{python}
oPs5ptC("+s_34(⟨2|3|1+2|4|1]-⟨2|4|1+2|3|1])mt²Δ_12|34|5\
(1/4⟨2|4⟩⟨3|4|2]-1/12⟨3|4|1]⟨1|4⟩-1/3⟨2|3|2]⟨3|4⟩)/(⟨1|2⟩[1|2]Δ_12|3|4|5²)")
<< mpc(real='-915.14788...', imag='525.36037...')
\end{minted}
The phase space point is pseudo-random, determined by the value of
\texttt{seed}, consequently for the same seed on the same machine and
python version it will return the same output. This may or may not
match the one displayed here.

It is also possible to generate a phase space point over a finite
field by changing the \texttt{Field} specifying the new characteristic
($(2^{31}-19)$ in the example) and number of digits (1). In the
\texttt{\_singular\_variety} function we now specify to how many
digits the constraints have to be satisfied (their
\textit{valuation}):
\begin{minted}[escapeinside=||, mathescape=true, linenos=False, numbersep=5pt, gobble=2, frame=lines, framesep=2mm, breaklines, breakautoindent=false, breakindent=-12.5pt]{python}
# finite-field phase space point
oPs8ptFp = Particles(8, field=Field("finite field", 2 ** 31 - 19, 1), seed=0)
oPs8ptFp._singular_variety(("⟨34⟩+[34]", "⟨34⟩-⟨56⟩", "⟨56⟩+[56]"),
                           (1, 1, 1), seed=0)
oPs8ptFp.mt = oPs8ptFp.m_t = - oPs8ptFp("⟨34⟩")
oPs8ptFp.mh = oPs8ptFp.m_h = oPs8ptFp.field.sqrt(oPs8ptFp("s_78"))
oPs5ptFp = oPs8ptFp.cluster([[1, ], [2, ], [3, 4], [5, 6], [7, 8]],
                            massive_fermions=((3, 'u', 1), (4, 'd', 1)))
\end{minted}
The evaluation of the same expression as the previous example over this finite field yields:
\begin{minted}[escapeinside=||, mathescape=true, linenos=False, numbersep=5pt, gobble=2, frame=lines, framesep=2mm, breaklines, breakautoindent=false, breakindent=-12.5pt]{python}
oPs5ptFp("+s_34(⟨2|3|1+2|4|1]-⟨2|4|1+2|3|1])mt²Δ_12|34|5\
(1/4⟨2|4⟩⟨3|4|2]-1/12⟨3|4|1]⟨1|4⟩-1/3⟨2|3|2]⟨3|4⟩)/(⟨1|2⟩[1|2]Δ_12|3|4|5²)")
<< 500540426 % 2147483629
\end{minted}

Finally we can generate a $p\kern0.2mm$-adic phase space point, with the prime
number $p=(2^{31}-19)$ and 10 digits of precision:
\begin{minted}[escapeinside=||, mathescape=true, linenos=False, numbersep=5pt, gobble=2, frame=lines, framesep=2mm, breaklines, breakautoindent=false, breakindent=-12.5pt]{python}
# 10-digit p-adic phase space point
oPs8ptQp = Particles(8, field=Field("padic", 2 ** 31 - 19, 10), seed=0)
oPs8ptQp._singular_variety(("⟨34⟩+[34]", "⟨34⟩-⟨56⟩", "⟨56⟩+[56]"),
                           (10, 10, 10), seed=0)
oPs8ptQp.mt = oPs8ptQp.m_t = - oPs8ptQp("⟨34⟩")
oPs8ptQp.mh = oPs8ptQp.m_h = oPs8ptQp.field.sqrt(oPs8ptQp("s_78"))
oPs5ptQp = oPs8ptQp.cluster([[1, ], [2, ], [3, 4], [5, 6], [7, 8]],
                            massive_fermions=((3, 'u', 1), (4, 'd', 1)))
\end{minted}
The evaluation of the same expression gives:
\begin{minted}[escapeinside=||, mathescape=true, linenos=False, numbersep=5pt, gobble=2, frame=lines, framesep=2mm, breaklines, breakautoindent=false, breakindent=-12.5pt]{python}
oPs5ptQp("+s_34(⟨2|3|1+2|4|1]-⟨2|4|1+2|3|1])mt²Δ_12|34|5\
(1/4⟨2|4⟩⟨3|4|2]-1/12⟨3|4|1]⟨1|4⟩-1/3⟨2|3|2]⟨3|4⟩)/(⟨1|2⟩[1|2]Δ_12|3|4|5²)")
<< 1776438348 + 1597211944*2147483629 + ... + O(2147483629^10)
\end{minted}

We also demonstrate the evaluation of a coefficient function
($c_{12\x34\x mmm}$ of eq.~\eqref{eq:c12x34xmmm}) at the three phase
space points over the different fields
\begin{minted}[escapeinside=||, mathescape=true, linenos=False, numbersep=5pt, gobble=2, frame=lines, framesep=2mm, breaklines, breakautoindent=false, breakindent=-12.5pt]{python}
from antares.terms.terms import Terms
oTri12x34xmmm = Terms("""
+s_34(⟨2|3|1+2|4|1]-⟨2|4|1+2|3|1])mt²Δ_12|34|5(1/4⟨2|4⟩⟨3|4|2]\
-1/12⟨3|4|1]⟨1|4⟩-1/3⟨2|3|2]⟨3|4⟩)/(⟨1|2⟩[1|2]Δ_12|3|4|5²)
...
+(-3/2mt⟨2|5|1]s_5tr(1+2|3+4)(⟨3|1+2|4]+[3|1+2|4⟩))/(Δ_12|34|5²)
""")
oTri12x34xmmm(oPs5ptC)
<< mpc(real='-84.27041...', imag='297.94955...')
oTri12x34xmmm(oPs5ptFp)
<< 181243021 % 2147483629
oTri12x34xmmm(oPs5ptQp)
<< 724832325 + 1137519716*2147483629 + ... + O(2147483629^10)
\end{minted}
where we have used ellipsis for some of the larger inputs and outputs,
and for simplicity we have hardcoded a string representation of this
coefficient in the notebook. The class \texttt{Terms} from
\textsc{Antares} \cite{antares} represents a rational function and is
based on a (quotient-)ring-agnostic implementation of monomials and
polynomials in \textsc{Syngular} \cite{syngular}. In particular, the
variables need not be defined, as long as a function capable of
evaluating them is provided, here in the form of the three instances
of the class \texttt{Particles} from \textsc{Lips}.

Then, an example is provided for the construction of a
$p\kern0.2mm$-adic phase space point in the limit
$Δ_{12|\three\four|\five}\rightarrow 0$ and the evaluation of the
previous coefficient at this phase space point
\begin{minted}[escapeinside=||, mathescape=true, linenos=False, numbersep=5pt, gobble=2, frame=lines, framesep=2mm, breaklines, breakautoindent=false, breakindent=-12.5pt]{python}
from antares.terms.terms import Terms
oPs8ptQpLimit = Particles(8, field=Field("padic", 2 ** 31 - 19, 10), seed=0)
oPs8ptQpLimit._singular_variety(("⟨34⟩+[34]", "⟨34⟩-⟨56⟩", "⟨56⟩+[56]",
                                "Δ_12|3456|78"), (10, 10, 10, 1, ), seed=0)
oPs8ptQpLimit.mt = - oPs8ptQpLimit("⟨34⟩")
oPs5ptQpLimit = oPs8ptQpLimit.cluster([[1,], [2,], [3, 4], [5, 6], [7, 8]],
                               massive_fermions=((3, 'u', 1), (4, 'd', 1)))
oTri12x34xmmm(oPs5ptQpLimit)
<< 1057813644*2147483629^-2 + ... + O(2147483629^7)
\end{minted}
The series now starts at $p^{-2}$ since this coefficient has a double
pole in $Δ_{12|\three\four|\five}$. The first three constraints are
satisfied to the full working precision of 10 digits, while the extra
constraint defining the kinematic limit is solved to one digit.

Lastly, we demonstrate the evaluation of all result coefficients in
the three number fields (the package \textsc{antares-results} can be
installed with \texttt{pip} or taken from GitHub)
\begin{minted}[escapeinside=||, mathescape=true, linenos=False, numbersep=5pt, gobble=2, frame=lines, framesep=2mm, breaklines, breakautoindent=false, breakindent=-12.5pt]{python}
from antares_results.ttH.qqttH.pm import coeffs
{coeff_name: coeff(oPs5ptC) for coeff_name, coeff in coeffs.items()}
<< {'tree': mpc(real='-0.24924...', imag='-0.19989...'),
<<  ...
<<  'bub12xmm': mpc(real='-0.09360...', imag='-0.00539...')}
{coeff_name: coeff(oPs5ptFp) for coeff_name, coeff in coeffs.items()}
<< {'tree': 941101849 % 2147483629,
<<  ...
<<  'bub12xmm': 582759043 % 2147483629}
{coeff_name: coeff(oPs5ptQp) for coeff_name, coeff in coeffs.items()}
<< {'tree': 123525663 + 918538244*2147483629 + ... + O(2147483629^10),
<<  ...
<<  'bub12xmm': 723645467 + 1821375353*2147483629 + ... + O(2147483629^10)}
\end{minted}
For the moment, we leave further exploration to the reader and recall
that Python’s object introspection allows for examining properties and
methods of objects at runtime.

\section{Scalar integrals}
\subsection{Loop integral definitions}
\label{Integrals}
We work in the Bjorken-Drell metric so that
$l^2=l_0^2-l_1^2-l_2^2-l_3^2$. The affine momenta $q_i$
are given by sums of the external momenta, $p_i$, 
where $q_n\equiv \sum_{i=1}^n p_i$ and $q_0 = 0$.
The propagator denominators are defined as $ d_i=(l+q_i)^2-m_{i+1}^2+i\varepsilon$.
The definition of the relevant scalar integrals is as follows,
\begin{eqnarray} \label{eq:scalarintegrals}
&&
  A_0(m_1)  =
 \frac{\mu^{4-D}}{i \pi^{\frac{D}{2}} \cG }\int \,
 \frac{d^D l} {d_0}\, , \nn \\
&&  B_0(p_1;m_1,m_2)  =
 \frac{\mu^{4-D}}{i \pi^{\frac{D}{2}}\cG }\int \,
 \frac{d^D l} {d_0 \; d_1}\, , \nn \\
&& C_0(p_1,p_2;m_1,m_2,m_3)  =
\frac{\mu^{4-D}}{i \pi^{\frac{D}{2}}\cG}
\int \, 
 \frac{d^D l}{d_0 \; d_1 \; d_2}\, ,\nn \\
&&D_0(p_1,p_2,p_3;m_1,m_2,m_3,m_4)
= \frac{\mu^{4-D}}{i \pi^{\frac{D}{2}}\cG }\int \, \frac{d^D l} {d_0 \; d_1 \; d_2\; d_3}\, ,\nn \\
&&E_0(p_1,p_2,p_3,p_4;m_1,m_2,m_3,m_4,m_5)
= \frac{\mu^{4-D}}{i \pi^{\frac{D}{2}} \cG}\int \,\frac{d^D l}{d_0 \; d_1 \; d_2\; d_3\; d_4}\, .
\end{eqnarray}
For the purposes of this paper we take the masses in the
propagators to be real.  Near four dimensions we use $D=4-2 \e$.  (For
clarity the small imaginary part which fixes the analytic
continuations is specified by $+i\,\varepsilon$).  $\mu$ is a scale introduced so that the integrals
preserve their natural dimensions, despite excursions away from $D=4$.
We have removed the overall constant which occurs in $D$-dimensional integrals,
\beq
\cG\equiv\frac{\Gamma^2(1-\e)\Gamma(1+\e)}{\Gamma(1-2\e)} = 
\frac{1}{\Gamma(1-\e)} +{\cal O}(\e^3) =
1-\e \gamma_E+\e^2\Big[\frac{\gamma_E^2}{2}-\frac{\pi^2}{12}\Big]
+{\cal O}(\e^3)\,.
\eeq
We have that
\beqn
B_0(p_3,0,m)&=&\frac{1}{\epsilon}+2+\ln\frac{\mu^2}{m^2}+O(\epsilon),\;\;\;p_3^2=m^2\, , \nn \\
A_0(m)&=&m^2 \Big(\frac{1}{\epsilon}+1+\ln\frac{\mu^2}{m^2}\Big)+O(\epsilon)\, .
\eeqn
We eliminate the tadpole integrals in favour of the bubble, $B_0(p_3,0,m)$ and a rational part,
using the identity,
\beq \label{eq:tadpole_elim}
A_0(m)=m^2\big(B_0(p_3,0,m)-1\big) + O(\epsilon)\, .
\eeq

\subsection{Reduction coefficients for $E_0(p_3,p_1,p_2,p_4,m,0,0,0,m)$}
\label{pentagon_reduction}
General reduction formulas to reduce scalar pentagon integrals to scalar box integrals
have been given in refs.~\cite{Melrose:1965kb,vanNeerven:1983vr,Bern:1993kr}.
For our case, where there are vanishing and equal masses we can derive simpler relations
for the real coefficients, $e_i$,
\begin{eqnarray} \label{fivetofour}
  E_0(p_3,p_1,p_2,p_4,m,0,0,0,m)&=& e_1 D_0(p_1,p_2,p_4,0,0,0,m) \nonumber \\
  &+&e_2 D_0(p_{13},p_2,p_4,m,0,0,m) \nonumber \\
  &+&e_3 D_0(p_3,p_{12},p_4,m,0,0,m) \nonumber \\
  &+&e_4 D_0(p_3,p_1,p_{24},m,0,0,m) \nonumber \\
  &+&e_5 D_0(p_3,p_1,p_2,m,0,0,0)\, ,
\end{eqnarray}
with $p_1^2=p_2^2=0,p_3^2=p_4^2=m^2,p_5=m_h^2$.
The coefficients are,
\begin{equation}
e_1=\frac{1}{2}\Big(
  \frac{\spba1.\four.2}{\spbaba1.\three.{(1+2)}.\four.2}  + {\rm c.c}
\Big)\, ,
\end{equation}
\begin{equation}
e_2=\frac{1}{2}\Big(
  \frac{\spaba2.\five.\four.2}{\spa2.1 \spbaba1.\three.(1+2).\four.2} + {\rm c.c}
\Big)\, ,
\end{equation}
\begin{eqnarray} 
  e_3&=&\frac{1}{2}\Big(
  \Big\{\frac{m^2 \spabab2.\four.(1+2).\four.2+\spaba2.\three.\four.1 \spbab1.\three.\four.2}{\spabab1.\three.(1+2).\four.2\spabab2.\four.(1+2).\three.1}
   +\frac{\spaba1.\three.\four.1}{\spabab1.\three.(1+2).\four.2\spa2.1} \nn \\
&-&\frac{
     \spaba1.\three.\four.2 \spba2.\three.2(\spab1.\four.1+\spab2.\four.2)}{\spa1.2\spabab1.\three.(1+2).\four.2\spabab2.\four.(1+2).\three.1}\Big\} + \Big\{{\rm c.c}\Big\}\Big)\, ,
\end{eqnarray}
\begin{equation}
e_4=\frac{1}{2}\Big(
  \frac{\spaba1.\five.\three.1}{\spa1.2 \spbaba2.\four.(1+2).\three.1} + {\rm c.c}
\Big)\, ,
\end{equation}
\begin{equation}
e_5=\frac{1}{2}\Big(
  \frac{\spba2.\three.1}{\spbaba2.\four.(1+2).\three.1} + {\rm c.c}
\Big)\, .
\end{equation}

The complex conjugation (c.c) is performed by exchanging all angle and square brackets.
An alternative form for $e_3$ which separates the complicated denominators at the cost
of introducing a spurious pole in $\spab1.\three.1$ is,
\begin{eqnarray} 
  e_3&=&\frac{1}{2} \Big(\Big\{\frac{m^2\spab2.\four.1}{\spabab2.\four.(1+2).3.1 \spab1.3.1}
  +\frac{\spaba1.3.\four.1}{\spabab1.3.(1+2).\four.2 \spa2.1}  \nn \\
  &+&\frac{\spab1.3.2\spaba1.3.\four.2}{\spab1.3.1\spabab1.3.(1+2).\four.2 \spa2.1}\Big\}+\Big\{ {\rm c.c.} \Big\}\Big)\,.
\end{eqnarray}
\bibliography{AnalyticRecon.bib}
\bibliographystyle{JHEP}
\end{document}